\documentclass[
twocolumn,
]{aastex631}

\usepackage{amsmath,amssymb}
\usepackage{graphicx} 
\usepackage{bm} 
\usepackage{makecell} 
\usepackage{threeparttable}
\usepackage{multirow} 
\usepackage{booktabs}
 
\usepackage[T1]{fontenc} 
\usepackage{color}

\begin{document}

\title{The origin channels of hierarchical binary black hole mergers in the LIGO-Virgo-KAGRA O1, O2, and O3 runs} 

\correspondingauthor{Xi-Long Fan}
\email{xilong.fan@whu.edu.cn}

\author{Guo-Peng Li}
\affiliation{
Department of Astronomy, School of Physics and Technology, Wuhan University, Wuhan 430072, China}

\author{Xi-Long Fan}
\affiliation{
Department of Astronomy, School of Physics and Technology, Wuhan University, Wuhan 430072, China}
\date{\today}

\begin{abstract}

We infer the origin channels of hierarchical mergers observed in the LIGO-Virgo-KAGRA (LVK) O1, O2, and O3 runs using a hierarchical Bayesian analysis under a parametric population model. By assuming the active galactic nucleus (AGN) disk and nuclear star cluster (NSC) channels, we find that NSCs likely dominate the hierarchical merger rate in the Universe, corresponding to a fraction of $f_{\rm NSC}=0.87_{-0.29}^{+0.10}$ at 90\% credible intervals in our fiducial model; AGN disks may contribute up to nearly half of hierarchical mergers detectable with LVK, specifically $f_{\rm det,AGN}=0.34_{-0.26}^{+0.38}$. We investigate the impact of the escape speed, along with other population parameters on the branching fraction, suggesting that the mass, mass ratio, and spin of the sources play significant roles in population analysis. We show that hierarchical mergers constitute at least $\sim$$10\%$ of the gravitational wave events detected by LVK during the O1-O3 runs. Furthermore, we demonstrate that it is challenging to effectively infer detailed information about the host environment based solely on the distribution of black hole merger parameters if multiple formation channels are considered.

\end{abstract}


\section{Introduction} \label{sec:intro}

During the first three observing runs, the LIGO~\citep{2015CQGra..32g4001L}, Virgo~\citep{2015CQGra..32b4001A}, and KAGRA~\citep{2012CQGra..29l4007S,2013PhRvD..88d3007A} (LVK) Collaboration reported nearly a hundred binary compact merger events~\citep{2019PhRvX...9c1040A,2021PhRvX..11b1053A,2021arXiv210801045T,2023PhRvX..13d1039A}. The majority of binary black hole (BBH) mergers could be explained by first-generation (1G) mergers formed from the collapse of stars. These primarily arise from isolated binary evolution, as well as dynamical formation in young/open star clusters and globular clusters~\citep{2021hgwa.bookE..16M,2022PhR...955....1M}. Others might stem from second (or higher) generation mergers, referred to hierarchical (repeated) mergers~\citep{2016ApJ...824L..12O,2017PhRvD..95l4046G,2017ApJ...840L..24F}, which could mainly originate from active galactic nucleus (AGN) disks and nuclear star clusters (NSCs)~\citep{2021NatAs...5..749G,2023Univ....9..138A}, although
hierarchical mergers can also occur in globular clusters (GCs) and young massive clusters (YSCs). Because they are less frequent in the latter environments, which strongly depends on the escape speed of their dynamical environments (see e.g.,~\citealp{2020PhRvD.102l3016A,2019PhRvD.100d3027R,2021MNRAS.505..339M,2021ApJ...915L..35K,2023MNRAS.520.5259A}).
Hierarchical mergers can efficiently bridge the pair-instability (PI) mass gap predicted by PI supernovae~\citep{2003ApJ...591..288H} and Pulsational PI supernovae~\citep{2007Natur.450..390W,2016A&A...594A..97B}, explaining the growth of intermediate-mass black holes (BHs) in dense stellar environments (e.g.,~\citealp{2020MNRAS.498..495D,2020MNRAS.497.1043D,2021ApJ...908..194T,2022ApJ...927..231F,2022ApJ...940..131G,2023PhRvD.108h3012K,2023A&A...673A...8D}).

In the active phase of galactic nuclei, a high-density gas disk can efficiently capture stellar-mass BHs and facilitate binary formation and mergers~\citep{2016ApJ...819L..17B,2017ApJ...835..165B,2020ApJ...898...25T}. This unique environment leads to distinctive characteristics of BBHs. The BH mass spectrum is expected to undergo hardening due to the orbital alignment of BHs with the AGN disks, which preferentially selects heavier BHs~\citep{2019ApJ...876..122Y}. Under accretion, the spin magnitudes of BHs are likely to be higher, with spin directions tending to align or anti-align with the disk~\citep{2007ApJ...661L.147B,2020ApJ...899...26T,2019ApJ...884L..12Y,2020MNRAS.494.1203M}. Moreover, migration and migration traps efficiently promotes BH hierarchical growth with extreme mass ratio mergers~\citep{2019PhRvL.123r1101Y,2022PhRvD.105f3006L}.

NSCs are also efficient sites for the occurrence of hierarchical mergers~\citep{2021NatAs...5..749G}. 
BBHs in NSCs are formed through dynamical pairing, either via exchanges between a binary star and an intruder, or via dynamical encounters of three initially single bodies~\citep{1975MNRAS.173..729H,1980AJ.....85.1281H}. The remnant of a 1G BBH merger can be retained by the host cluster if the escape speed of the host cluster is larger than its kick velocity. This surviving remnant can then pair up with another BH to form a hierarchical BBH. Therefore, the crucial condition for hierarchical mergers in NSCs is the escape speed of the host cluster---higher escape speeds result in more efficient hierarchical mergers~\citep{2019PhRvD.100d1301G,2022A&A...666A.194L,2023PhRvD.107f3007L}. Additionally, the occurrence of relatively symmetric binary for hierarchical mergers may be preferential in NSCs due to mass segregation~\citep{2022MNRAS.515.1830P}.

As a result, the difference (e.g., the mass, mass ratio, and spin of binaries) in hierarchical mergers lies in the presence of additional changes in the mass and spin of BHs in AGN disks compared to those in NSCs.
This implies that the mass and effective spin distributions of hierarchical mergers in AGN disks are higher than those in NSCs~\citep{2023PhRvD.107f3007L}. Regarding the mass ratios of hierarchical mergers, unequal-mass pairings are favored in AGN disks, represented by the hierarchical branch of \texttt{NG+1G} mergers (\texttt{NG} refers to the BH generation), while in NSCs, relatively equal-mass pairings are still favored, represented by the hierarchical branch of \texttt{NG+NG} mergers. In addition, it is generally agreed that the efficiency of BH hierarchical growth in AGN disks may be higher than in NSCs (e.g.,~\citealp{2019PhRvL.123r1101Y,2023PhRvD.107f3007L}), although it depends on several parameters that are currently unknown. These parameters include the formation channel of NSCs, the real structure and dynamics of an AGN disk, the possible existence of migration traps, the presence of a supermassive BH in the NSC, and the dynamical evolution of the NSC and its co-evolution with the surrounding environment~\citep{2021NatAs...5..749G,2023Univ....9..138A,2023ApJ...944L..42L}.

Primarily focusing on 1G mergers, previous research has examined the branching fractions of their isolated binary evolution or cluster channels for BBH mergers~\citep{2021ApJ...910..152Z,2023ApJ...955..127C,2023arXiv231217491I,2024A&A...685A..51V}. However, some of the BBH mergers detected by LVK might be hierarchical mergers originating from dynamical channels, such as AGN disks and NSCs. In this study, our objective is to infer the branching fractions of the AGN disk and NSC channels for hierarchical mergers. Our focus is on investigating the population parameters~(see Table\,\ref{tab:model}) that influence branching fractions, which suggests that effectively inferring information about the host environment based solely on the distribution of BBH merger parameters presents a significant challenge.

The rest of this paper is organized as follows. In Section~\ref{sec:method}, we provide a detailed description of the methodology used in this study, including hierarchical Bayesian analysis in Section~\ref{subsec:Bayes}, hierarchical merger models in Section~\ref{subsec:model}, and hierarchical merger events in Section~\ref{subsec:event}. In Section~\ref{sec:result}, we present the main results of our analysis, highlighting the key findings and their implications. In Section~\ref{sec:cd}, we summarize our conclusions and 
provide a discussion of these results, including potential uncertainties, comparisons with previous work, and the broader implications for the field.

\section{Method} \label{sec:method}

We employ a hierarchical Bayesian analysis to explore the origins of hierarchical mergers in the first three observing runs. We simulate hierarchical mergers with a parametric population model and apply numerical relativity fits to determine each merger remnant’s total mass~\citep{2012ApJ...758...63B}, final spin~\citep{2016ApJ...825L..19H}, and kick velocity~\citep{2007ApJ...659L...5C}.

\subsection{Hierarchical Bayesian analysis} \label{subsec:Bayes}

To infer the parameters describing our population model, we employ a hierarchical Bayesian approach using a set of data $\{d_i\}$ from $N_{\rm det}$ GW detections, for which the Bayesian formalism is applied with the likelihood~\citep{2019MNRAS.486.1086M,2019PASA...36...10T}
\begin{equation}\label{likelihood}
\begin{aligned}
    \mathcal{L}(\{d_i\}|\Lambda,\{\mu_j\})
    \propto
    \prod \limits_{i=1}^{N_{\rm det}}
    \frac{\int {\rm d}\theta\,\mathcal{L}(d_i|\theta)\,\mathcal{L}(\theta|\Lambda,\{\mu_j\})}
    {\xi(\Lambda,\{\mu_j\})}\,,
\end{aligned}
\end{equation}
where $\Lambda=\{f_j\}$ is the branching fraction of the $j$-th origin channel with a population model $\mu_j$ with $V_{{\rm esc},j}$ the escape speed of the host; 
$\theta=(\mathcal{M},\chi_{{\rm eff}},q)$ is a set of parameters with the chirp mass, effective spin, and mass ratio, representing the characteristics of individual events; $\xi(\Lambda,\{\mu_j\})$ is the fraction of hierarchical mergers that are detectable for a population model. Note that there are further BBH merger parameters that also can play a crucial role in determining the actual origin of GW sources, such as redshift~\citep{2023Univ....9..138A}, which for simplicity, we do not consider additional parameters and will do so in the future as more gravitational wave events occur.

We consider the conditional prior (i.e., population model) $\mathcal{L}(\theta|\Lambda,\{\mu_j\})$ for hierarchical mergers is contributed by the AGN disk channel and the NSC channel:
\begin{equation}
\begin{aligned}
    \mathcal{L}(\theta|\Lambda,\{\mu_j\})=
    \sum\limits_{j=1}^{2} f_j\,\mathcal{L}(\theta|\mu_j)\,,
\end{aligned}
\end{equation}
where $f_1+f_2=1$. The detectable fraction is 
\begin{equation}
\xi(\Lambda,\{\mu_j\}) = \sum\limits_{j=1}^{2} f_j\,\xi(\mu_j)\,,
\end{equation}
and correspondingly the detectable branching fraction $f_{{\rm det},j} = \xi(\mu_j)\,f_j / \xi(\Lambda,\{\mu_j\})$. 
To calculate $\xi(\mu_j)$, we assume that redshifts of the simulated mergers are drawn uniformly in comoving volume between $z\in[0,2]$, and that the generated gravitational waves conform to PhemonA~\citep{2007CQGra..24S.689A}. Then, we can calculate the signal-to-noise ratio (SNR) according to $\rho^2 = \frac{16}{5} \int \frac{(2fT)S_{\rm h}(f)}{S_{\rm n}(f)}d(\ln f)$, where $f$ is frequency of the gravitational wave, $T$ is the observation time, $S_{\rm h}(f)$ is the one-sided, averaged, power spectral density of the signal, and $S_{\rm n}(f)$ is the noise sensitivity curve of LIGO~\citep{2020LRR....23....3A}. When SNR reaches the predetermined threshold, we consider the signal to be detectable~\citep{Abadie-2010-Abbott-CQGra..27q3001A}.

The likelihood $\mathcal{L}(d_i|\theta)$ is in terms of the posterior probability density function (PDF) $p(\theta_i|d_i)$, which can be computed by using some default prior $\pi(\theta_i|\varnothing)$. In practice, the posterior PDF $p(\theta_i|d_i)$ can be discretely sampled with $S_i$ samples from the posterior, $\{\theta_{i,k}\}$, for $k \in [1,S_i]$, where the posterior PDFs of GW events we use are from~\citet{2023ApJ...946...59N}. Then, one can replace the integral in Eq.\,(\ref{likelihood}) with a discrete sum over PDF samples:
\begin{equation}\label{sl}
\begin{aligned}
\mathcal{L}(\{d_i\}|\Lambda,\{\mu_j\})
\propto
\prod \limits_{i=1}^{N_{\rm det}}
\frac{1}{\xi(\Lambda,\{\mu_j\})}
\sum\limits_{j=1}^{2}
\frac{f_j}{S_i}
\sum\limits_{k=1}^{S_i}
\frac{\mathcal{L}(\theta_{i,k}|\mu_j)}
{\pi(\theta_{i,k}|\varnothing)}\,.
\end{aligned}
\end{equation}

\subsection{Hierarchical merger models} \label{subsec:model}

\begin{table*}[t]
\centering
\caption{\label{tab:model}%
Summary of the models.}
\resizebox{\linewidth}{!}{
\begin{tabular}{lcccccccccccc}
\hline
Model & \multicolumn{3}{c}{Mass} & \multicolumn{3}{c}{Spin} & \multicolumn{2}{c}{Mass ratio} & \multicolumn{1}{c}{Escape speed} & \multicolumn{2}{c}{Branch} & \multicolumn{1}{c}{SNR} \\
\cline{2-13} 
& $\alpha_{\rm m,AGN}$ & $\alpha_{\rm m,NSC}$ & $m_{\rm max}\,[M_\odot$] & $\alpha_{\rm \chi,AGN}$ & $\beta_{\rm \chi,AGN}$ & $\gamma_{\rm \chi,AGN}$  & $\beta_{\rm q,AGN}$& $\beta_{\rm q,NSC}$ & $V_{\rm esc,NSC}\,[{\rm km\,s^{-1}}]$ & AGN & NSC & - \\
\hline
\texttt{1, Fiducial}          & 1.0 & 2.3 & 80 & 1.5 & 3.0 & 1.0  & 0.0  & 5.0 & 100 & \texttt{NG+1G}       & \texttt{NG+NG}        & $>$$8$  \\
\texttt{Model\,2}             & 2.0 & 3.5 & 80 & 1.5 & 3.0 & 1.0  & 0.0  & 5.0 & 100 & \texttt{NG+1G}       & \texttt{NG+NG}        & $>$$8$  \\
\texttt{Model\,3}             & 1.0 & 2.3 & 65 & 1.5 & 3.0 & 1.0  & 0.0  & 5.0 & 100 & \texttt{NG+1G}       & \texttt{NG+NG}        & $>$$8$  \\
\texttt{Model\,4}             & 1.0 & 2.3 & 80 & 2.0 & 2.5 & 1.0  & 0.0  & 5.0 & 100 & \texttt{NG+1G}       & \texttt{NG+NG}        & $>$$8$  \\
\texttt{Model\,5}             & 1.0 & 2.3 & 80 & 1.5 & 3.0 & 0.5  & 0.0  & 5.0 & 100 & \texttt{NG+1G}       & \texttt{NG+NG}        & $>$$8$  \\
\texttt{Model\,6}             & 1.0 & 2.3 & 80 & 1.5 & 3.0 & 2.0  & 0.0  & 5.0 & 100 & \texttt{NG+1G}       & \texttt{NG+NG}        & $>$$8$  \\
\texttt{Model\,7}             & 1.0 & 2.3 & 80 & 1.5 & 3.0 & 1.0  & -1.0 & 1.0 & 100 & \texttt{NG+1G}       & \texttt{NG+NG}        & $>$$8$  \\
\texttt{Model\,8}             & 1.0 & 2.3 & 80 & 1.5 & 3.0 & 1.0  & 0.0  & 5.0 & 50  & \texttt{NG+1G}       & \texttt{NG+NG}        & $>$$8$  \\
\texttt{Model\,9}             & 1.0 & 2.3 & 80 & 1.5 & 3.0 & 1.0  & 0.0  & 5.0 & 200 & \texttt{NG+1G}       & \texttt{NG+NG}        & $>$$8$  \\
\texttt{Model\,10}            & 1.0 & 2.3 & 80 & 1.5 & 3.0 & 1.0  & 0.0  & 5.0 & 300 & \texttt{NG+1G}       & \texttt{NG+NG}        & $>$$8$  \\
\texttt{Model\,11}            & 1.0 & 2.3 & 80 & 1.5 & 3.0 & 1.0  & 0.0  & 5.0 & 500 & \texttt{NG+1G}       & \texttt{NG+NG}        & $>$$8$  \\
\texttt{Model\,12}            & 1.0 & 2.3 & 80 & 1.5 & 3.0 & 1.0  & 0.0  & 5.0 & 100 & \texttt{NG+1G}       & \texttt{NG+$\leq$NG}  & $>$$8$  \\
\texttt{Model\,13}            & 1.0 & 2.3 & 80 & 1.5 & 3.0 & 1.0  & 0.0  & 5.0 & 100 & \texttt{NG+$\leq$NG} & \texttt{NG+NG}        & $>$$8$  \\
\texttt{Model\,14}            & 1.0 & 2.3 & 80 & 1.5 & 3.0 & 1.0  & 0.0  & 5.0 & 100 & \texttt{NG+$\leq$NG} & \texttt{NG+$\leq$NG}  & $>$$8$  \\
\texttt{Model\,15}            & 1.0 & 2.3 & 80 & 1.5 & 3.0 & 1.0  & 0.0  & 5.0 & 100 & \texttt{NG+1G}       & \texttt{NG+NG}        & $>$$12$ \\
\hline
\end{tabular}}
\begin{tablenotes} 
\item {\bf Note.}
\textit{Column 1} (Model): Name of the model. In particular, \texttt{Model\,1} is presented as a fiducial model.
\textit{Columns 2,\,3,\,4} (Mass): $\alpha_{\rm m,AGN}$ and $\alpha_{\rm m,NSC}$ are the spectral indices for the power law of the \textsc{PowerLaw+Peak} model in AGN disks and NSCs, respectively; $m_{\rm max}$ is the maximum mass of 1G BH mass distribution. 
\textit{Columns 5,\,6,\,7} (Spin): $\alpha_{\rm \chi,AGN}$ and $\beta_{\rm \chi,AGN}$ are the standard shape parameters of the beta distribution for spins in AGN disks; $\gamma_{\rm \chi,AGN}$ is the index that describes the misalignment angle distribution between the BH spin and the orbital angular momentum of AGN disks. 
\textit{Columns 8,\,9} (Mass ratio): $\beta_{\rm q,AGN}$ and $\beta_{\rm q,NSC}$ are the mass ratio spectral indices in AGN disks and NSCs, respectively.
\textit{Column 10} (Escape speed): $V_{\rm esc,NSC}$ is the escape speed of NSCs.
\textit{Columns 11,\,12} (Branch): Hierarchical merger branches in AGN disks and NSCs. 
\textit{Column 13} (SNR): The SNR threshold. 
\end{tablenotes} 
\end{table*}

We employ a parametric population model described in~\citet{2023PhRvD.107f3007L} enabling us to rapidly simulate hierarchical mergers in both AGN disks and NSCs across diverse population parameter variations. This approach, or similar variations, has already been used in several works to interpret some gravitational-wave (GW) sources with peculiar features that may hint at a hierarchical merger origin (see e.g.,~\citealp{2021ApJ...915L..35K,2022arXiv220905766M}), which indicates the feasibility of using this method to explore hierarchical mergers currently.
Particularly, our model employs the same numerical code across all considered scenarios, enabling a direct comparison of the merging properties of different channels without the potential biases introduced by using different simulation codes for BBH catalogs. In this framework, the two dynamical BBH formation channels are differentiated by their initial BH distributions in mass and spin, pairing probabilities, escape velocities, and hierarchical merger branches. Below, we provide a detailed guide on how to assign the corresponding parameter values for NSCs and AGN disks, as summarized in Table~\ref{tab:model}.

The masses of 1G BHs follow a distribution described by the \textsc{PowerLaw+Peak} model~\citep{2023PhRvX..13a1048A}. In this model, two key parameters govern the mass distribution: the spectral index $\alpha_{\rm m}$ for the power law and maximum mass $m_{\rm max}$ of 1G BHs. Notably, it has been extensively shown that BHs in the PI mass gap can be produced via stellar collisions and interactions~\citep{2024MNRAS.534.1634L}; as the most massive in the sample, these BHs have a large probability of merging with other BHs (see e.g.,~\citealp{2020ApJ...903...45K,2021MNRAS.507.5132D,2021MNRAS.501.5257R,2022MNRAS.512..884R,2024MNRAS.528.5140A}), and can be mistakenly interpreted as higher-generation mergers when they are, in fact, 1G mergers~\citep{2024arXiv240619044A}. Therefore, we analyze branching fractions for different values of $\alpha_{\rm m}$ and $m_{\rm max}$, with other parameters in the \textsc{PowerLaw+Peak} model derived from~\citep{2023PhRvX..13a1048A}. In the \texttt{Fiducial} model, we set $\alpha_{\rm m,AGN}=1.0$, consistent with AGN disk hardening effects on the BH mass function~\citep{2019ApJ...876..122Y}. These studies assumed an initial power-law mass distribution and found that the index steepens by approximately $\Delta\alpha_{\rm m}\sim1.3$, resulting in a more top-heavy population of merging BHs in AGN disks. Correspondingly for nuclear star clusters (NSCs), we set $\alpha_{\rm m,NSC}=2.3$, aligning with the Kroupa initial mass function~\citep{2001MNRAS.322..231K} in cases without gas hardening. In \texttt{Model\,2}, we adopt $\alpha_{\rm m,AGN}=2.0$ and $\alpha_{\rm m,NSC}=3.5$, based on the LVK Collaboration's inferred $\alpha=3.5^{+0.6}_{-0.56}$ in GWTC-3~\citep{2023PhRvX..13a1048A}.  Additionally, we set $m_{\rm max} = 80\,M_\odot$ and $65\,M_\odot$, corresponding to the absence and presence of the PI mass gap in the \texttt{Fiducial} model and \texttt{Model\,3}, respectively.


The spin magnitudes of 1G BHs are uniformly assigned within a beta distribution~\citep{2023PhRvX..13a1048A}, determined by the standard shape parameters $\alpha_{\rm \chi}$ and $\beta_{\rm \chi}$. In NSCs, spin magnitudes are expected to be small, so we assign them according to the results of~\citep{2023PhRvX..13a1048A}; the spin tilt angles are randomly selected from an isotropic distribution across a sphere. In contrast, in AGN disks, spin magnitudes are likely to be higher due to angular momentum exchange with the disks~\citep{2019ApJ...884L..12Y,2020MNRAS.494.1203M,2020ApJ...899...26T}. Therefore, we assume spin magnitudes in AGN disks to be larger than in NSCs, as adopted in the \texttt{Fiducial} model and \texttt{Model\,4}. For spin directions in AGN disk, we assume a distribution for the misalignment angle $\theta$ between the BH spin and the orbital angular momentum of AGN disks, given by $p(\theta) \propto |\theta-\pi/2|^{\gamma_{\rm \chi}}$. This assumption is made due to the torque exerted by gas accretion from AGN disks, which tends to align or anti-align the spin with $\theta\to0~{\rm or}~\pi$~\citep{2007ApJ...661L.147B,2020ApJ...899...26T,2019ApJ...884L..12Y,2020MNRAS.494.1203M}. As a comparison, we use $\gamma_{\rm \chi,AGN}=1.0$, $0.5$, and $2.0$ in the \texttt{Fiducial} model, \texttt{Model\,5}, and \texttt{Model\,6}, respectively.


The mass ratios of 1G BBHs are drawn from the distribution $p(q) \propto q^{\beta_{\rm q}}$, where $q=m_2/m_1$ represents the ratio between the secondary BH mass ($m_2$) and the primary BH mass ($m_1$) in the binary system~\citep{2023PhRvX..13a1048A}. In the \texttt{Fiducial} model, we set $\beta_{\rm q,AGN}=0.0$ and $\beta_{\rm q,NSC}=5.0$, which is motivated by runaway mergers occurring due to migration and migration traps in AGN disks~\citep{2016ApJ...819L..17B,2020MNRAS.494.1203M}, and by mass segregation in NSCs~\citep{2019PhRvD.100d3027R}. For the control group, represented by \texttt{Model\,7}, we adopt $\beta_{\rm q,AGN}=-1.0$ and $\beta_{\rm q,NSC}=1.0$, where a value of $\sim$1.0 aligns with the analysis result by the LVK Collaboration~\citep{2023PhRvX..13a1048A}.

Then, we can pair 1G BBHs and synthesize hierarchical mergers if the remnants of the 1G mergers can be retained by the host environment, with the escape speed $V_{\rm esc}$ exceeding the kick velocity $V_{\rm kick}$. For NSCs, the escape speed generally ranges from $\sim$$10$ to $600\,{\rm km\,s^{-1}}$~\citep{2016ApJ...831..187A}, with a typical escape speed of $\sim$$100\,{\rm km\,s^{-1}}$ adopted in the \texttt{Fiducial} model. Models \texttt{Model\,8} to \texttt{Model\,11} are used for comparison. In AGNs, we neglect the kicks of merger remnants~\citep{2020ApJ...901L..34Y,2020ApJ...898...25T,2022PhRvD.105f3006L} due to the high orbital velocities, $\sim$$2\times10^4\,{\rm km\,s^{-1}}$, and the small kick magnitudes, as BH spins are typically aligned or anti-aligned with the disk~\citep{2020MNRAS.494.1203M}.

To simulate a complete hierarchical merger population, we iterate until 6G progenitor BHs are formed, capturing over $99\%$ of hierarchical mergers. Note that the NSC in our model includes a supermassive BH~\citep{2022MNRAS.517.5827F}. We do not account for the impact of stellar binaries, as their contribution is not expected to significantly alter our branching fraction estimates. In the densest and most massive star clusters, a substantial fraction of stellar binaries are typically ionized before they can form a BBH~\citep{2015ApJ...800....9M,2016ApJ...831..187A}.

There are three hierarchical BBH merger branches: \texttt{NG+1G}, \texttt{NG+NG}, and \texttt{NG+$\leq$NG}~\citep{2022ApJ...935L..20Z,2023PhRvD.107f3007L}. In particular, \texttt{NG+1G} mergers are expected to occur preferentially in AGNs because of migration traps~\citep{2012MNRAS.425..460M,2016ApJ...819L..17B}; \texttt{NG+NG} mergers tend to occur in star clusters because of mass segregation~\citep{2022ApJ...935L..20Z}; while \texttt{NG+$\leq$NG} mergers, which include but are not limited to \texttt{NG+1G} and \texttt{NG+NG}, represent a steady-state limit. 
For \texttt{NG+$\leq$NG} mergers, we pair each $N$G BH with another BH with the generation $M$ ($M\leq N$). The probability of the generation $M$ obeys $p(M)\propto 2^{-(M-1)}$. For example, a $N$G BH is twice as likely to merge with a 1G BH than a 2G BH, and four times as likely to merge with a 1G BH than a 3G BH.
In the case of hierarchical mergers, pairing may crucially depend on binary formation timing and the BH-burning process, where dynamical relaxation ejects the most massive BHs from the cluster center. These processes can impact even second-generation mergers, as relaxation timescales in NSCs are relatively long ($>$$10^9$\,yr)~\citep{2012ApJ...759...52D}. The assumption that nearly equal-mass binaries are more common in NSCs due to mass segregation may not be entirely accurate, as self-consistent simulations largely show that forming multiple high-generation BH chains is highly suppressed~\citep{2010ApJ...715.1006K}. Here, we investigate branching fractions under various reasonable combinations of hierarchical branches. In the \texttt{Fiducial} model, we assume that \texttt{NG+1G} and \texttt{NG+NG} mergers dominate hierarchical merger rates in AGNs and star clusters, respectively, based on the efficiency of migration traps and mass segregation. Alternative reasonable combinations of hierarchical branches are explored in \texttt{Model\,12}, \texttt{Model\,13}, and \texttt{Model\,14}.

In addition, selection effects is one of the key ingredients in (GW) population analyses, and it is essential to estimate these effects as accurately as possible. To compute this accurately, one must consider multiple aspects, including the various detectors active during observing runs and their duty cycles, changes in detector sensitivity across and within observing runs, noise distributions in the detectors, and distributions of all source parameters, including those not included in population models. All of these ingredients impact detectability. The standard approach is to rerun the same searches that detected real GW candidates on simulated signals by injecting fake GW signals into simulated detector noise and then checking which signals are recovered by the search pipelines~(see e.g.,~\citealp{KAGRA:2021duu}). 
For convenience, we do not implement the standard approach as described above. Instead, we use a simplified method (see Section~\ref{subsec:Bayes}). This approach allows us to maintain consistent and reliable outcomes while simplifying the computational process. To ensure our estimates are unbiased and not limited by an approximate detectable fraction, we set thresholds of SNR~$>$$8$ and SNR~$>$$12$ in the \texttt{Fiducial} model and \texttt{Model\,15}, respectively, to calculate detectable fractions for comparison.

We note that for a population-level analysis, one should measure the joint posterior distribution of all population parameters simultaneously. However, investigating the relevant parameter space of hierarchical mergers is computationally challenging. As a result, in this work, we focus on model comparison to investigate the impact of population parameters on the branching fraction, i.e., computing the likelihood in Eq.\,(\ref{sl}) at a fixed set of values $\mu_j$, and then inferring the posterior distribution of branching fractions.

In summary, we construct 15 different models, each capable of independently synthesizing hierarchical mergers in both NSCs and AGN disks. It is important to note that the population parameters are numerous, such as those in the \textsc{PowerLaw+Peak} mass distribution model, which also includes additional parameters not fully detailed here. For instance, there includes $m_{\rm min}$ the minimum mass of the power-law component of the primary mass distribution, $\lambda_{\rm peak}$ the fraction of BBH systems in the Gaussian component, $\mu_m$ the mean of the Gaussian component in the primary mass distribution, $\sigma_m$ the width of the Gaussian component in the primary mass distribution, and $\delta_n$ the range of mass tapering at the lower end of the mass distribution (see Table VI in~\citealp{2023PhRvX..13a1048A}). The parameters listed in Table~\ref{tab:model} are the key ones we focus on in this study. Parameters not included in Table~\ref{tab:model} are kept constant in our analysis, and their specific values can be found in ~\citet{2023PhRvX..13a1048A} and~\citet{2023PhRvD.107f3007L}. In the models presented in Table~\ref{tab:model}, \texttt{Model\,1} serves as our fiducial  model. \texttt{Model\,2} and \texttt{Model\,3} explore the impact of changes in the mass distribution on the branching fraction. \texttt{Model\,4}, \texttt{Model\,5} and \texttt{Model\,6} investigate how changes in the spin distribution affect the branching fraction. \texttt{Model\,7} examine the influence of changes in the mass ratio distribution. Models from \texttt{Model\,8} and \texttt{Model\,11} explore the impact of changes in escape speed of the host. Models from \texttt{Model\,12} and \texttt{Model\,14} asses how variations in hierarchical branch impact the branching fraction, while \texttt{Model\,15} is to ensure our estimates are unbiased due to the simple calculation for election effects.

\subsection{Hierarchical merger events} \label{subsec:event}

\begin{table*}[t]
\centering
\caption{\label{tab:HM}%
Summary of hierarchical candidate events ($p_{\rm hier}>0.5$).}
\resizebox{\linewidth}{!}{
\begin{tabular}{lccccccccccccccccc}
\hline
No. &Name & \texttt{Fiducial}& \texttt{Model\,2}& \texttt{Model\,3}& \texttt{Model\,4}& \texttt{Model\,5}& \texttt{Model\,6}& \texttt{Model\,7}& \texttt{Model\,8}& \texttt{Model\,9}& \texttt{Model\,10}& \texttt{Model\,11}& \texttt{Model\,12}& \texttt{Model\,13}& \texttt{Model\,14}& \texttt{Model\,15} \\
\hline
1 & GW151012\_095443 &  & \checkmark &  & \checkmark &  &  &  &  &  &  &  &  &  &  &  \\ 
2 & GW170202\_135657 & \checkmark & \checkmark & \checkmark & \checkmark & \checkmark & \checkmark &  & \checkmark & \checkmark & \checkmark &  & \checkmark & \checkmark & \checkmark & \checkmark \\ 
3 & \textbf{GW170729\_185629} & \checkmark & \checkmark & \checkmark & \checkmark & \checkmark & \checkmark & \checkmark & \checkmark & \checkmark & \checkmark & \checkmark & \checkmark & \checkmark & \checkmark & \checkmark \\ 
4 & GW190404\_142514 &  & \checkmark &  &  &  &  &  &  &  &  &  &  &  &  &  \\ 
5 & GW190412\_053044 & \checkmark & \checkmark & \checkmark & \checkmark & \checkmark & \checkmark &  & \checkmark & \checkmark & \checkmark & \checkmark & \checkmark & \checkmark & \checkmark & \checkmark \\ 
6 & GW190512\_180714 &  & \checkmark &  &  &  &  &  &  &  &  &  &  &  &  &  \\ 
7 & \textbf{GW190517\_055101} & \checkmark & \checkmark & \checkmark & \checkmark & \checkmark & \checkmark & \checkmark & \checkmark & \checkmark & \checkmark & \checkmark & \checkmark & \checkmark & \checkmark & \checkmark \\ 
8 & \textbf{GW190519\_153544} & \checkmark & \checkmark & \checkmark & \checkmark & \checkmark & \checkmark & \checkmark & \checkmark & \checkmark & \checkmark & \checkmark & \checkmark & \checkmark & \checkmark & \checkmark \\ 
9 & \textbf{GW190521\_030229} & \checkmark & \checkmark & \checkmark & \checkmark & \checkmark & \checkmark & \checkmark & \checkmark & \checkmark & \checkmark & \checkmark & \checkmark & \checkmark & \checkmark & \checkmark \\ 
10 & GW190602\_175927 &  & \checkmark & \checkmark & \checkmark &  &  &  &  & \checkmark & \checkmark & \checkmark &  &  &  &  \\ 
11 & \textbf{GW190620\_030421} & \checkmark & \checkmark & \checkmark & \checkmark & \checkmark & \checkmark & \checkmark & \checkmark & \checkmark & \checkmark & \checkmark & \checkmark & \checkmark & \checkmark & \checkmark \\ 
12 & \textbf{GW190706\_222641} & \checkmark & \checkmark & \checkmark & \checkmark & \checkmark & \checkmark & \checkmark & \checkmark & \checkmark & \checkmark & \checkmark & \checkmark & \checkmark & \checkmark & \checkmark \\ 
13 & GW190708\_232457 &  & \checkmark &  &  &  &  &  &  &  &  &  &  &  &  &  \\ 
14 & GW190719\_215514 &  & \checkmark & \checkmark & \checkmark &  &  &  &  & \checkmark &  & \checkmark &  &  &  &  \\ 
15 & \textbf{GW190814\_211039} & \checkmark & \checkmark & \checkmark & \checkmark & \checkmark & \checkmark & \checkmark & \checkmark & \checkmark & \checkmark & \checkmark & \checkmark & \checkmark & \checkmark & \checkmark \\ 
16 & GW190828\_065509 &  & \checkmark &  & \checkmark &  &  &  &  &  &  &  &  &  &  &  \\ 
17 & GW190916\_200658 &  &  &  &  &  &  &  &  &  &  & \checkmark &  &  &  &  \\ 
18 & \textbf{GW190929\_012149 }& \checkmark & \checkmark & \checkmark & \checkmark & \checkmark & \checkmark & \checkmark & \checkmark & \checkmark & \checkmark & \checkmark & \checkmark & \checkmark & \checkmark & \checkmark \\ 
19 & GW191109\_010717 &  &  & \checkmark &  &  &  & \checkmark &  &  &  & \checkmark &  &  &  &  \\ 
20 & \textbf{GW191127\_050227} & \checkmark & \checkmark & \checkmark & \checkmark & \checkmark & \checkmark & \checkmark & \checkmark & \checkmark & \checkmark & \checkmark & \checkmark & \checkmark & \checkmark & \checkmark \\ 
21 & GW200115\_042309 &  & \checkmark &  &  &  &  &  &  &  &  & \checkmark &  &  &  &  \\ 
22 & \textbf{GW200129\_114245} & \checkmark & \checkmark & \checkmark & \checkmark & \checkmark & \checkmark & \checkmark & \checkmark & \checkmark & \checkmark & \checkmark & \checkmark & \checkmark & \checkmark & \checkmark \\ 
23 & GW200216\_220804 &  & \checkmark &  &  &  &  &  &  &  &  & \checkmark &  &  &  &  \\ 
24 & GW200225\_060421 &  & \checkmark &  &  &  &  &  &  &  &  &  &  &  &  &  \\ 
25 & GW200306\_093714 & \checkmark & \checkmark & \checkmark & \checkmark & \checkmark & \checkmark & \checkmark & \checkmark & \checkmark & \checkmark &  & \checkmark & \checkmark & \checkmark & \checkmark \\ 
\multicolumn{2}{l}{Total Number} & 13 & 23 & 16 & 17 & 13 & 13 & 12 & 13 & 15 & 14 & 17 & 13 & 13 & 13 & 13\\
\hline
\end{tabular}}
\begin{tablenotes} 
\item {\bf Note.}
Bold events indicate that they are hierarchical candidate events in all different models.
\end{tablenotes} 
\end{table*}

In general, the properties of hierarchical merger events include massive components, high spins, significantly asymmetric masses, etc.~\citep{2021NatAs...5..749G}. To perform this type of population analyses self consistently, one should include all detected events over the observation window and incorporate models for all formation channels. Therefore, we introduce an additional channel for 1G (isolated) BBHs, which does not involve any hierarchical component. This channel encompasses isolated binary evolution, GCs, YSCs, and other formation pathways~\citep{2021hgwa.bookE..16M,2022PhR...955....1M}. In this channel, we strictly constrain BBHs based on the analysis results from the GWTC-3 by the LVK Collaboration~\citep{2023PhRvX..13a1048A}. In total, there are five BBH merger populations, consisting of three 1G merger populations and two hierarchical merger populations. The 1G populations originate from NSCs, AGN disks, and the extra 1G channel we introduced. The probability that the $i$-th GW event originates from the $n$-th 1G channel ($m=1$) or the $n$-th hierarchical channel ($m=2$) is given by
\begin{equation}
\begin{aligned}
p(d_i|\mu_{m,n}) = 
\frac{1}{S_i}
\sum\limits_{k=1}^{S_i}
\frac{\mathcal{L}(\theta_{i,k}|\mu_{m,n})}
{\pi(\theta_{i,k}|\varnothing)}\,,
\end{aligned}
\end{equation}
where $\mu_{m,n}$ is the population model, and $n=1,2,3$ for $m=1$ and $n=1,2$ for $m=2$. The probabilities of a GW event originating from a 1G merger or a hierarchical merger are
\begin{equation}\label{eq1h}
\begin{aligned}
p(d_i|{\rm 1G}) &= \sum\limits_{n=1}^{3}p(d_i|\mu_{1,n})\,,\\
p_(d_i|{\rm hier}) &= \sum\limits_{n=1}^{2}p(d_i|\mu_{2,n})\,.
\end{aligned}
\end{equation}
The normalized probability of a GW event originating from a hierarchical merger is given by 
\begin{equation}
\begin{aligned}
p_{\rm hier}(d_i) = \frac{p(d_i|{\rm hier})}{p(d_i|{\rm 1G}) + p(d_i|{\rm hier})}\,.
\end{aligned}
\end{equation}
We consider a GW event as a potential hierarchical candidate if $p_{\rm hier} > 0.5$.

As a result, we find that approximately 12 to 23 GW events are potential hierarchical candidates, as listed in Table\,\ref{tab:HM}. In particular, ten GW events are identified as hierarchical candidates across all models, encompassing almost all the hierarchical merger events proposed in previous studies (e.g.,~\citealp{2021ApJ...915L..35K,2022PhRvD.106j3013M,2023PhRvD.108h4044A,2024arXiv241102778C,2022ApJ...941L..39W,2024PhRvL.133e1401L}). This suggests the reliability of our method, despite not carefully considering all possible channels. Table\,\ref{tab:HM} shows that hierarchical mergers make up $\sim$$10\%\text{-}25\%$ of the GW events detected by LVK during the O1-O3 observing runs.

\section{Results} \label{sec:result}

We infer $\Lambda = \{f_j\} = \{f_{\rm AGN}, f_{\rm NSC}\}$ by assuming a uniform prior, where $f_{\rm AGN} + f_{\rm NSC} = 1$. These represent the branching fractions in AGN disks and NSCs. Therefore, we only need to estimate the posterior distribution of one parameter, such as $f_{\rm AGN}$, which is done by varying the population model parameter set $\{\mu_j\}$, as outlined in Table~\ref{tab:model}.

\subsection{Channel contribution}\label{subsec:cc}

\begin{figure*}
\centering
\includegraphics[width=16cm]{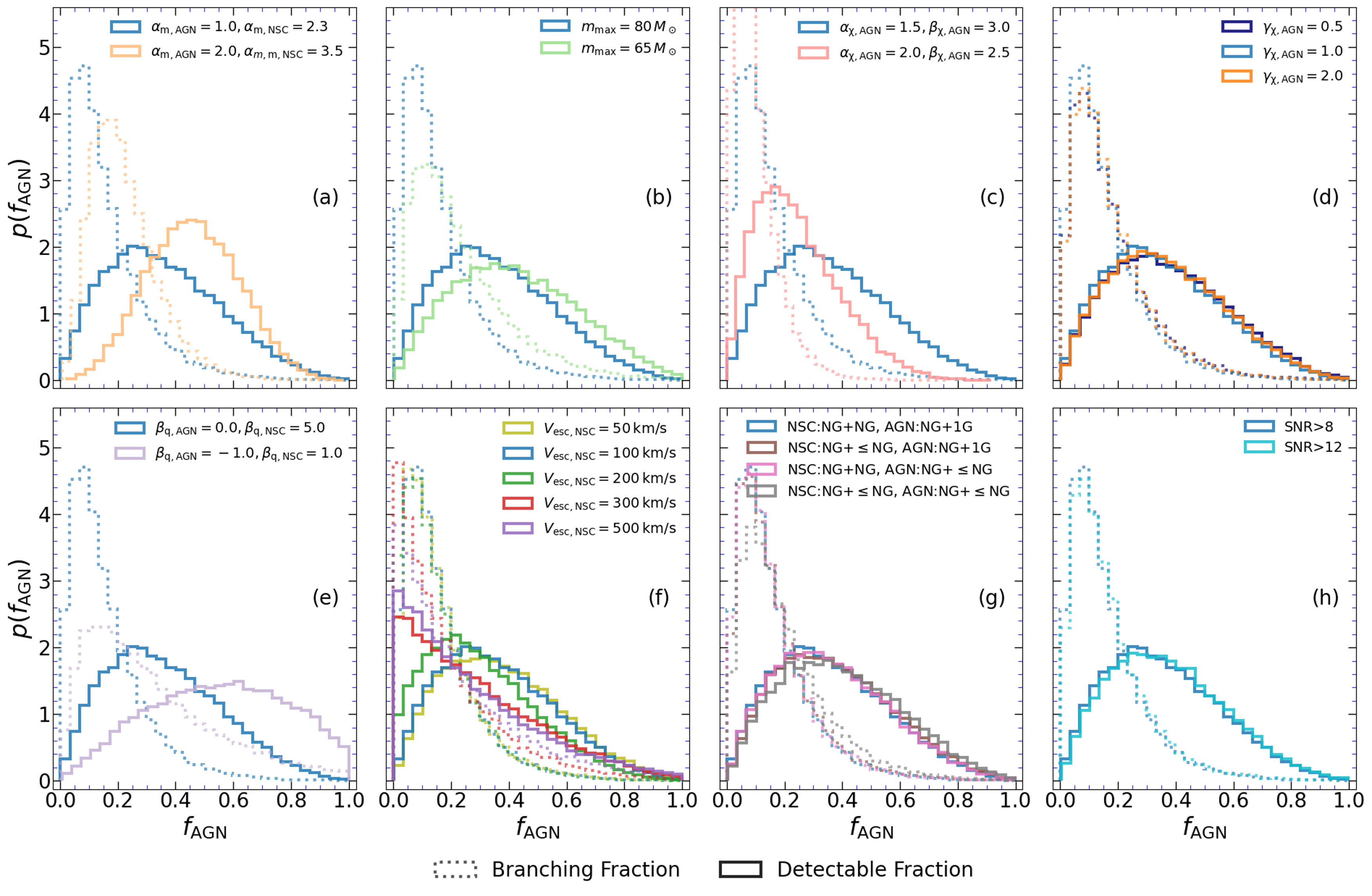}
\caption{
The probability distributions of the branching fraction (dashed) and detectable fraction (solid) for hierarchical mergers in the AGN disk channel are shown under the different population models in Table~\ref{tab:model}. The results are divided into 8 subgraphs, with each subgraph representing a class of comparable models 
(except for the parameters listed in the legend of each subgraph, all other parameters remain unchanged). Each subgraph includes a blue legend, representing the analysis result from the \texttt{Fiducial} model for comparison. Since different subgraphs display different parameter variations, the content of the blue legend differs in each case.
}\label{fig:1} 
\end{figure*}

Figure\,\ref{fig:1} compares the posterior distributions of the branching fraction and detectable fraction for the different population models listed in Table~\ref{tab:model}. Across all models, we find that the NSC channel dominates the occurrence of hierarchical mergers over the majority of the population parameter space corresponding to $1-f_{\rm AGN}$ being always much larger than 0.5, consistent with the predictions for the result (or rate) presented in~\citet{2022MNRAS.517.5827F}. We also observe that the AGN disk channel could contribute up to nearly half of hierarchical mergers detectable by LVK, particularly for \texttt{Model\,7} (see Fig.\,\ref{fig:1}\,(e)), which favors extremely asymmetric binaries in AGN disks. It should be noted that, in previous work, we mistakenly concluded that most of the hierarchical merger events detected by LVK might originate from the AGN channel~\citep{2023PhRvD.107f3007L}.

From Fig.\,\ref{fig:1}, parameters with a significant impact on the branching fractions include $\alpha_{\rm m}$, $\alpha_{\rm \chi}$, $\beta_{\rm \chi}$, and $\beta_{\rm q}$ (see Fig.\,\ref{fig:1}\,(a,\,c,\,\&\,e)), all of which strongly indicate that the mass, mass ratio, and spin of BBH mergers play important roles in population analysis. 
In particular:
i) The mass index, $\alpha_{\rm m}$, determines the slope of the 1G BH mass distribution. A smaller $\alpha_{\rm m}$ corresponds to a more top-heavy population of merging BHs. Detailed analysis reveals a positive correlation between $f_{\rm AGN}$ and $\Delta\alpha_{\rm m} = |\alpha_{\rm m,NSC} - \alpha_{\rm m,AGN}|$. This suggests that a relatively harder mass spectral index in one channel leads to a greater contribution from that channel, as hierarchical mergers tend to involve more massive components (see Fig.\,\ref{fig:1}\,(a)).
ii) The standard shape parameters, $\alpha_{\rm \chi}$ and $\beta_{\rm \chi}$, define the spin magnitudes of 1G BHs within a beta distribution. For AGN disks, $\alpha_{\rm \chi,AGN}=2.0$ and $\beta_{\rm \chi,AGN}=2.5$ represent higher spin magnitudes compared to $\alpha_{\rm \chi,AGN}=1.5$ and $\beta_{\rm \chi,AGN}=3.0$. These higher spin magnitudes result in a lower contribution of AGN disks to the hierarchical merger rate (see Fig.\,\ref{fig:1}\,(c)).
iii) The mass-ratio index, $\beta_{\rm q}$, determines the slope of the 1G BBH mass ratio distribution. Fig.\,\ref{fig:1}\,(e) shows that hierarchical mergers in AGN disks favor asymmetric masses. With $\beta_{\rm q,AGN}=-1.0$ favoring more asymmetric binaries, the detectable fraction of hierarchical mergers in AGN disks becomes dominant among the hierarchical mergers observed by LVK.

Parameters with a minor impact on the branching fractions are $m_{\rm max}$ and $V_{\rm esc,NSC}$ (see Fig.\,\ref{fig:1}\,(b\,\&\,f)). The former is understandable because even if some 1G BHs lie in the PI mass gap, they represent a very small fraction; particularly, among nearly 100 GW sources detected by LVK, only a small number of events have masses in the PI mass gap. Regarding the escape speed, its minor impact arises from the interplay between two factors: while a larger escape speed promotes more effective hierarchical mergers, it also leads to a higher proportion of massive black holes, reducing the selection effects. These two factors cancel each other out, making the escape speed's impact less significant. Parameters with minimal impact include $\gamma_{\rm \chi,AGN}$ and the hierarchical merger branch (see Fig.\,\ref{fig:1}\,(d\,\&\,g)). Additionally, Fig.\,\ref{fig:1}\,(h) shows that our estimations are not biased by the use of a sub-standard approximate method for considering selection effects.

We observe that a low-mass distribution corresponding to a large spectral index $\alpha_{\rm m}$ tends to increase the branching fraction and detectable fraction for AGN disks, as shown in Fig.\,\ref{fig:1}\,(a). This is due to the fact that massive BHs are favored in hierarchical mergers, while the low-mass distribution in both AGN disks and NSCs reduces the gap in selection effects between the two channels. similarly, the distributions of spin and mass ratio of mergers are generally not related to selection effects, as reflected in the synchronous right or left shifts of the branching fraction and detectable fraction distributions in Fig.\,\ref{fig:1}\,(c\,\&\,e). Specifically, asymmetric binaries are favored in hierarchical mergers; however, the detectable fraction shows a relatively flat distribution under models with different mass ratio distributions. This is because, in contrast to 1G mergers, hierarchical mergers are more chaotic, leading to greater uncertainty. Figure\,\ref{fig:1}\,(c) shows that although hierarchical merging tends towards larger spins, it is not necessarily better to have larger spins.


\subsection{The relationship between branching fraction and escape speed}\label{subsec:re}

Although we show that the parameter for the escape speed of NSCs has a minor impact on the branching fraction (see Fig.\,\ref{fig:1}\,(f)), it is important to further explore its relationship, as escape speed is the parameter in our population model that directly reflects the host environment.

\begin{figure}
\centering
\includegraphics[width=8cm]{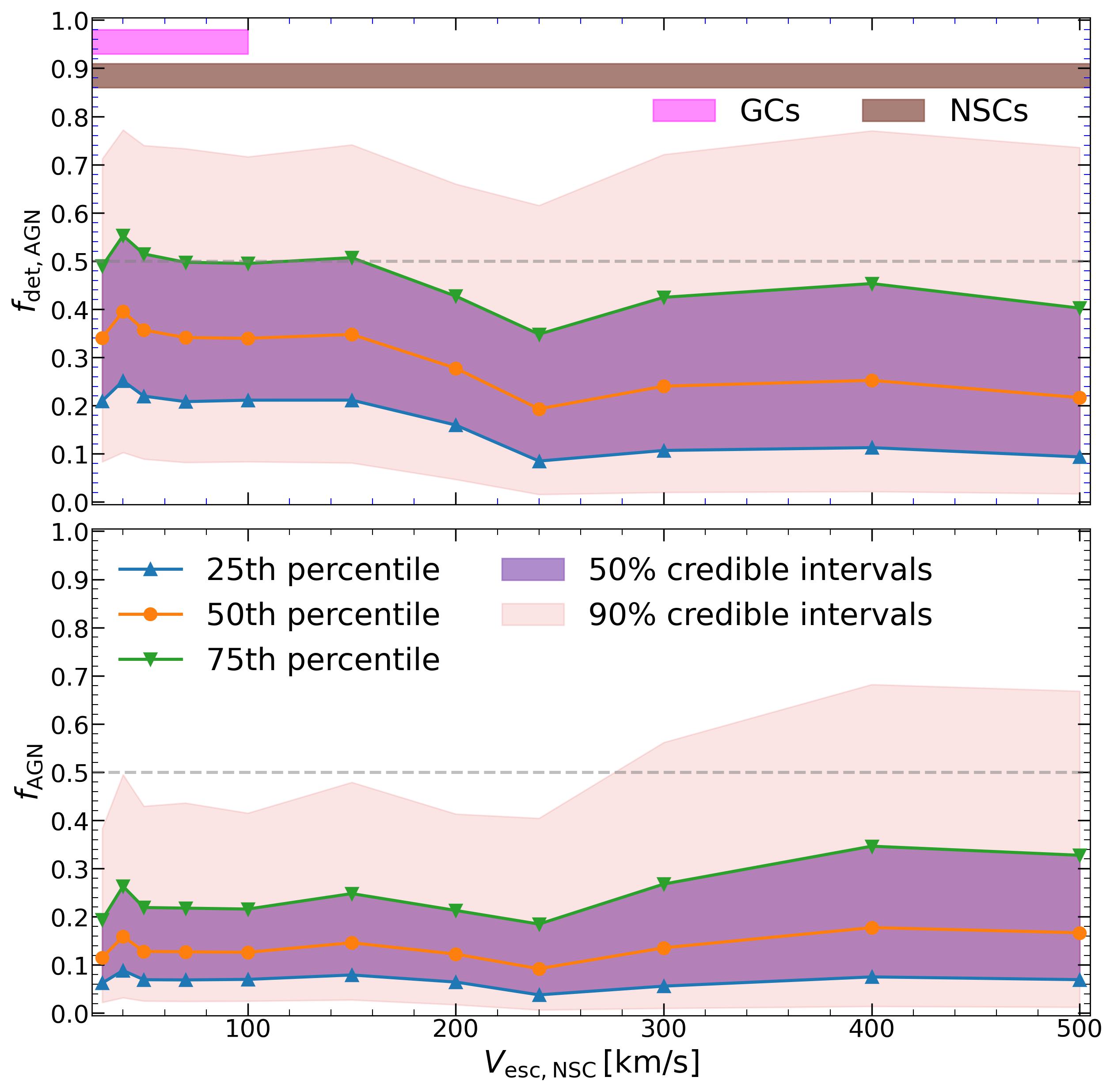}
\caption{
The escape speed of NSCs vs. the detectable fractions (\textit{top}) and branching fractions (\textit{bottom}) of the AGN disk channel are shown. The percentiles and confidence intervals of the distributions of the detectable and branching fractions for the AGN disk channel, as a function of the escape speed of NSCs, are plotted. We also show the known ranges for the escape velocities of GCs and NSCs for comparison.
}\label{fig:2}
\end{figure}

We analyze the relationship between the branching fraction and the escape speed of NSCs by expanding the model to vary the escape speed. In Fig.~\ref{fig:2}, we present the branching and detectable fractions at different escape speeds. For comparison, the escape speed of a typical GC is $\sim 2\text{-}100\,{\rm km\,s^{-1}}$, while that of an NSC ranges from $\sim 10\text{-}600\,{\rm km\,s^{-1}}$~\citep{2016ApJ...831..187A}. The trend suggests that as the escape speed increases, the distributions of the detectable fraction tend to decrease slightly, while the branching fraction increases slightly. The 50\% credible intervals of the distributions for both detectable fractions and branching fractions are always less than $\sim 0.5$ and $\sim 0.3$, respectively. Finally, from Figs.~\ref{fig:1} and \ref{fig:2}, the distributions of branching and detectable fractions show significant uncertainty. This underscores the challenge of effectively inferring detailed host environment information solely from the distribution of BH merger parameters, especially when considering multiple formation channels, such as the escape speed.

\section{Conclusions and Discussion} \label{sec:cd}

We constrain the origin channels of hierarchical binary black hole mergers in the LVK O1-O3 runs by employing a hierarchical Bayesian analysis. We find that hierarchical mergers constitute at least $\sim$10\% of the GW events detected by LVK during the O1-O3 runs. Following that, we suggest that the NSC channel would dominate the hierarchical merger rate, while AGN disks could contribute up to nearly half of hierarchical mergers detectable with LVK, Specifically, we obtain fractions of $f_{\rm NSC}=0.87_{-0.29}^{+0.10}$ and $f_{\rm det,AGN}=0.34_{-0.26}^{+0.38}$ at 90\% credible intervals in our fiducial model. The branching fraction to hierarchical mergers primarily hinges on the key population parameters, including the mass spectral index $\alpha_{\rm m}$, the standard shape parameters of the spin distribution $\alpha_{\rm \chi}$ and $\beta{\rm \chi}$, and the mass ratio spectral index $\beta_{\rm q}$. This suggests that the mass, mass ratio, and spin of BBH mergers play important roles in population analysis. Other population parameters have relatively minor impacts on the branching fraction.

Notably, most of the population parameters do not significantly affect the branching fraction~(see Fig.\,\ref{fig:1}), with only a few parameters showing considerable sensitivity, such as $\alpha_{\rm m}$, $\alpha_{\rm \chi}$ and $\beta{\rm \chi}$, and $\beta_{\rm q}$. Additionally, the uncertainty in the calculated branching fractions is large, which further complicates the extraction of information about the population parameters, or the host environment. In particular, the escape speed of NSCs has a minor effect on the branching fraction~(see Fig.\,\ref{fig:2}), even though the hierarchical growth efficiency of BHs in star clusters primarily depends on the escape speed. In conclusion, deducing detailed information about the host environment solely from the distribution of BH merger parameters becomes challenging when multiple formation channels are considered.

Recent studies have employed various approaches to constrain the contribution of BBH mergers from AGN disks, with a particular focus on hierarchical mergers driven by AGNs. Specifically, 
\citet{2023MNRAS.526.6031V} compared two scenarios: one in which AGNs are physically associated with BBH mergers, and another where the presence of AGNs in the localization volumes of GW events is purely coincidental. Their results showed that the most luminous AGNs do not contribute significantly to the majority of detected BBH mergers in the local Universe. This suggests that the connection between AGNs and BBH mergers is more intricate than previously assumed.
\citet{2024arXiv240813144L} pointed out that a significant fraction of BBHs in AGN disks may experience strong lensing by the central supermassive BH. Therefore, the absence of lensed GW signals could provide a way to constrain the fraction of BBH binaries in AGN disks. It has been suggested that, with the next-generation detectors and a large number of BBH observations, the lack of lensed events could help exclude certain scenarios as the primary sites for BBH mergers. 
They found that with non-detection of lensing with the current $\mathcal{O}(100)$ events from the LVK O1-O3 runs, the 90th percentile upper bound on the fraction of AGN-BBHs events is $\gtrsim$$50\%$ of all LVK binaries.
Our analysis suggests that $\sim$$34\%$ of the hierarchical mergers detected by LVK originate from AGN disks. Assuming hierarchical mergers constitute $\sim$$15\%$ of all LVK events, this corresponds to $\sim$$5\%$ of the total LVK mergers being attributed to AGN disks. This result aligns with the previous estimate of the contribution of BBH mergers from AGN disks~\citep{2024arXiv240813144L}.
For hierarchical mergers, \citet{2022ApJ...941L..39W} and \citet{2024PhRvL.133e1401L,2024ApJ...977...67L} have identified 
a sub-population in LIGO-Virgo BBH mergers, a fraction $\sim$$5\%$
of which may originate from AGN disks (the hierarchical mergers they identified are broadly consistent with our findings).
In addition, \citet{2021ApJ...918L..31M,2024ApJ...975..117M} proposed a Bayesian framework to predict the merger history focusing on hierarchical mergers involving previous mergers of the binary components, although they do not further predict a branching fraction.
Building on these studies, future research will likely continue refining our understanding of the role of AGN disks in BBH mergers, particularly through more precise observational data and advanced modeling techniques. As next-generation detectors become operational, they may provide stronger constraints on hierarchical mergers, further elucidating the contributions of the AGN-driven channel to the overall BBH merger rate.

Furthermore, in AGN disks, BBH mergers may associate with electromagnetic emissions~(e.g.,~\citealp{2019ApJ...884L..50M,2021ApJ...916L..17W,2024ApJ...961..206C,2023ApJ...955...23T,2023MNRAS.524.6015L,2024arXiv240610904Z}) or high-energy neutrinos~(e.g.,~\citealp{2023ApJ...955...23T,2023ApJ...951...74Z,2024MNRAS.528L..88Z}). Our result suggests that the AGN disk channel may contribute up to nearly half of hierarchical mergers detectable with LVK. This is expected to be constrained and verified through the spatial correlation between the sky maps of GW events and the positions of AGNs~\citep{2022MNRAS.514.2092V,2023MNRAS.526.6031V,2024arXiv240505318V,2024arXiv240721568V}. We expect the discovery of confident of GW events, particularly in AGN disks, as well as associated electromagnetic counterparts~\citep{2020PhRvL.124y1102G,2023ApJ...942...99G,2024arXiv240505318V}. 
The detection of multi-messenger signals, including both GWs and electromagnetic counterparts, will be crucial in providing a more comprehensive understanding of BBH mergers, especially in AGN disks. These multi-messenger observations will not only help confirm the contribution of AGN disks to hierarchical mergers but also offer insights into the astrophysical environments of these mergers. By correlating GW events with electromagnetic emissions or high-energy neutrinos, we can better understand the physical processes at play in AGN disks and improve our knowledge of the underlying merger mechanisms. Furthermore, such observations will provide an important test of general relativity in extreme environments and open new avenues for cosmological applications, such as precise measurements of the Hubble constant and the expansion rate of the Universe.
We anticipate that with the fourth and fifth observing runs of Advanced LIGO ~\citep{2015CQGra..32g4001L}, Advanced Virgo~\citep{2015CQGra..32b4001A}, and KAGRA~\citep{2012CQGra..29l4007S,2013PhRvD..88d3007A}, along with third-generation GW detectors in operation~\citep{2010CQGra..27s4002P,2010CQGra..27h4007P,2017CQGra..34d4001A}, the increasing data on GW events will enable us to constrain origin channels and their host environments more precisely.

Here, we comment that our results may have the following inaccuracies.
Although hierarchical mergers generally tend to involve massive components, high spins, and asymmetric masses~\citep{2021NatAs...5..749G}, a small portion deviates from this pattern~\citep{2023PhRvD.107f3007L}. This implies the presence of hidden hierarchical mergers within the 1G merger events that have not been selected. These mergers are more likely to originate from the NSC channel rather than the AGN disk channel. This might further reduce the importance of the AGN disk channel in GW BBH sources. 
By integrating the mass distributions of 1G and hierarchical mergers in Eq.\,(\ref{eq1h}) for masses below $10\,M_\odot$, we obtain an overall ratio of $\sim$$15:1$ for 1G to hierarchical mergers. Given the predominance of 1G mergers, this ratio is expected to increase further. Considering that the number of observed GW events in the LVK O1-O3 runs  with primary masses below $10\,M_\odot$ is around 10, this suggests that, at most, only one hierarchical merger event may be present within this sample. Consequently, this does not affect our conclusion that the NSC channel is the dominant site for hierarchical mergers, although this potential effect may warrant careful consideration when a significantly larger number of events are detected by next-generation detectors.

Our results are derived under the assumption that hierarchical mergers predominantly occur in AGN disks and NSCs, without considering the contribution of other channels such as GCs and YSCs~\citep{2021NatAs...5..749G}. If the contribution of these clusters to the hierarchical merger population is not neglected~\citep{2022A&A...666A.194L}, our results may not hold true. This concern is particularly relevant given the potential scenario where the number of BBH mergers originating from globular clusters exceeds those from NSCs and AGNs ~\citep{2022LRR....25....1M}.
This may result in a reduction in the overall contribution of NSC and AGN disk channels to our results, particularly for NSCs. This is because GCs and NSCs belong to a similar class of dynamical channels, with the primary distinction being the difference in escape speeds. The specific impact and contribution of these effects will require further investigation in the future. However, as shown in our analysis in Fig.\,\ref{fig:2}, when the escape speed is low, corresponding to a GC-like environment~\citep{2016ApJ...831..187A}, the contribution of the AGN channel remains largely unchanged. This demonstrates that at least the AGN disk channel would not play the primary role in hierarchical mergers.

A parametric population model, which does not fully incorporate the complexity of the physical processes governing the formation of high-generation mergers in AGN disks and NSCs, is employed in this study as an approximation of the detailed evolution of hierarchical mergers in dynamically rich environments. It may not capture important intricacies of hierarchical assembly, especially in environments that are gas-rich or involve interactions with a central super-massive BH. Given the inherent limitations of the current astrophysical models, the ultimate aim of more accurate inference with expanding such methodologies for hierarchical mergers should be kept in mind. Despite the use of a parametric population model in the hierarchical Bayesian analysis, we can gain an illustrative understanding of how branching fractions scale with escape speed and BBH merger information by varying the parameters of our hierarchical merger model.
One notable advantage of this approach is that the use of the same numerical code for different environments eliminates potential biases arising from variations in simulation frameworks. Additionally, the computational cost is relatively low, particularly when inferring only $f$, the branching fraction. Within the context of this model, we can also gain insights into key aspects of the system. For instance, we identify which population parameters (i.e., $\alpha_{\rm m}$, $\alpha_{\rm \chi}$ and $\beta{\rm \chi}$, and $\beta_{\rm q}$) play dominant roles in determining the branching fraction, while others have negligible influence. These findings suggest that the relative roles of NSCs and AGN disks in hierarchical mergers remain consistent, irrespective of minor parameter variations.

\section{Acknowledgments}
We would like to thank the referees for their valuable comments which considerably improved the original text. 
We are grateful to Xiang-Nan Shen, Xiao-Yan Li, and Huan Zhou for helpful suggestion. 
This work is supported by National Key R$\&$D Program of China (2020YFC2201400), the National Natural Science Foundation of China (grant No. 11922303), and the Fundamental Research Funds for the Central Universities, Wuhan University (grant No. 2042022kf1182).


\begin{thebibliography}{}
\expandafter\ifx\csname natexlab\endcsname\relax\def\natexlab#1{#1}\fi
\providecommand{\url}[1]{\href{#1}{#1}}
\providecommand{\dodoi}[1]{doi:~\href{http://doi.org/#1}{\nolinkurl{#1}}}
\providecommand{\doeprint}[1]{\href{http://ascl.net/#1}{\nolinkurl{http://ascl.net/#1}}}
\providecommand{\doarXiv}[1]{\href{https://arxiv.org/abs/#1}{\nolinkurl{https://arxiv.org/abs/#1}}}

\bibitem[{{Abadie} {et~al.}(2010){Abadie}, {Abbott}, {Abbott}, {Abernathy}, {Accadia}, {Acernese}, {Adams}, {Adhikari}, {Ajith}, {Allen}, \& et~al.}]{Abadie-2010-Abbott-CQGra..27q3001A}
{Abadie}, J., {Abbott}, B.~P., {Abbott}, R., {et~al.} 2010, Classical and Quantum Gravity, 27, 173001, \dodoi{10.1088/0264-9381/27/17/173001}

\bibitem[{{Abbott} {et~al.}(2017){Abbott}, {Abbott}, {Abbott}, {Abernathy}, {Ackley}, {Adams}, {Addesso}, {Adhikari}, {Adya}, {Affeldt}, \& et~al.}]{2017CQGra..34d4001A}
{Abbott}, B.~P., {Abbott}, R., {Abbott}, T.~D., {et~al.} 2017, Classical and Quantum Gravity, 34, 044001, \dodoi{10.1088/1361-6382/aa51f4}

\bibitem[{{Abbott} {et~al.}(2019){Abbott}, {Abbott}, {Abbott}, {Abraham}, {Acernese}, {Ackley}, {Adams}, {Adhikari}, {Adya}, {Affeldt}, \& et~al.}]{2019PhRvX...9c1040A}
---. 2019, Physical Review X, 9, 031040, \dodoi{10.1103/PhysRevX.9.031040}

\bibitem[{{Abbott} {et~al.}(2020){Abbott}, {Abbott}, {Abbott}, {Abraham}, {Acernese}, {Ackley}, {Adams}, {Adya}, {Affeldt}, {Agathos}, \& et~al.}]{2020LRR....23....3A}
---. 2020, Living Reviews in Relativity, 23, 3, \dodoi{10.1007/s41114-020-00026-9}

\bibitem[{{Abbott} {et~al.}(2021){Abbott}, {Abbott}, {Abraham}, {Acernese}, {Ackley}, {Adams}, {Adams}, {Adhikari}, {Adya}, {Affeldt}, \& et~al.}]{2021PhRvX..11b1053A}
{Abbott}, R., {Abbott}, T.~D., {Abraham}, S., {et~al.} 2021, Physical Review X, 11, 021053, \dodoi{10.1103/PhysRevX.11.021053}

\bibitem[{{Abbott} {et~al.}(2023{\natexlab{a}}){Abbott}, {Abbott}, {Acernese}, {Ackley}, {Adams}, {Adhikari}, {Adhikari}, {Adya}, {Affeldt}, {Agarwal}, \& et~al.}]{2023PhRvX..13d1039A}
{Abbott}, R., {Abbott}, T.~D., {Acernese}, F., {et~al.} 2023{\natexlab{a}}, Physical Review X, 13, 041039, \dodoi{10.1103/PhysRevX.13.041039}

\bibitem[{{Abbott} {et~al.}(2023{\natexlab{b}}){Abbott}, {Abbott}, {Acernese}, {Ackley}, {Adams}, {Adhikari}, {Adhikari}, {Adya}, {Affeldt}, {Agarwal}, \& et~al.}]{2023PhRvX..13a1048A}
---. 2023{\natexlab{b}}, Physical Review X, 13, 011048, \dodoi{10.1103/PhysRevX.13.011048}

\bibitem[{Abbott {et~al.}(2023)}]{KAGRA:2021duu}
Abbott, R., {et~al.} 2023, Phys. Rev. X, 13, 011048, \dodoi{10.1103/PhysRevX.13.011048}

\bibitem[{{Acernese} {et~al.}(2015){Acernese}, {Agathos}, {Agatsuma}, {Aisa}, {Allemandou}, {Allocca}, {Amarni}, {Astone}, {Balestri}, {Ballardin}, \& et~al.}]{2015CQGra..32b4001A}
{Acernese}, F., {Agathos}, M., {Agatsuma}, K., {et~al.} 2015, Classical and Quantum Gravity, 32, 024001, \dodoi{10.1088/0264-9381/32/2/024001}

\bibitem[{{Ajith} {et~al.}(2007){Ajith}, {Babak}, {Chen}, {Hewitson}, {Krishnan}, {Whelan}, {Br{\"u}gmann}, {Diener}, {Gonzalez}, {Hannam}, {Husa}, {Koppitz}, {Pollney}, {Rezzolla}, {Santamar{\'\i}a}, {Sintes}, {Sperhake}, \& {Thornburg}}]{2007CQGra..24S.689A}
{Ajith}, P., {Babak}, S., {Chen}, Y., {et~al.} 2007, Classical and Quantum Gravity, 24, S689, \dodoi{10.1088/0264-9381/24/19/S31}

\bibitem[{{Antonelli} {et~al.}(2023){Antonelli}, {Kritos}, {Ng}, {Cotesta}, \& {Berti}}]{2023PhRvD.108h4044A}
{Antonelli}, A., {Kritos}, K., {Ng}, K. K.~Y., {Cotesta}, R., \& {Berti}, E. 2023, \prd, 108, 084044, \dodoi{10.1103/PhysRevD.108.084044}

\bibitem[{{Antonini} \& {Gieles}(2020)}]{2020PhRvD.102l3016A}
{Antonini}, F., \& {Gieles}, M. 2020, \prd, 102, 123016, \dodoi{10.1103/PhysRevD.102.123016}

\bibitem[{{Antonini} \& {Rasio}(2016)}]{2016ApJ...831..187A}
{Antonini}, F., \& {Rasio}, F.~A. 2016, \apj, 831, 187, \dodoi{10.3847/0004-637X/831/2/187}

\bibitem[{{Antonini} {et~al.}(2024){Antonini}, {Romero-Shaw}, \& {Callister}}]{2024arXiv240619044A}
{Antonini}, F., {Romero-Shaw}, I.~M., \& {Callister}, T. 2024, arXiv e-prints, arXiv:2406.19044, \dodoi{10.48550/arXiv.2406.19044}

\bibitem[{{Arca sedda} {et~al.}(2024){Arca sedda}, {Kamlah}, {Spurzem}, {Rizzuto}, {Giersz}, {Naab}, \& {Berczik}}]{2024MNRAS.528.5140A}
{Arca sedda}, M., {Kamlah}, A. W.~H., {Spurzem}, R., {et~al.} 2024, \mnras, 528, 5140, \dodoi{10.1093/mnras/stad3951}

\bibitem[{{Arca Sedda} {et~al.}(2023{\natexlab{a}}){Arca Sedda}, {Mapelli}, {Benacquista}, \& {Spera}}]{2023MNRAS.520.5259A}
{Arca Sedda}, M., {Mapelli}, M., {Benacquista}, M., \& {Spera}, M. 2023{\natexlab{a}}, \mnras, 520, 5259, \dodoi{10.1093/mnras/stad331}

\bibitem[{{Arca Sedda} {et~al.}(2023{\natexlab{b}}){Arca Sedda}, {Naoz}, \& {Kocsis}}]{2023Univ....9..138A}
{Arca Sedda}, M., {Naoz}, S., \& {Kocsis}, B. 2023{\natexlab{b}}, Universe, 9, 138, \dodoi{10.3390/universe9030138}

\bibitem[{{Aso} {et~al.}(2013){Aso}, {Michimura}, {Somiya}, {Ando}, {Miyakawa}, {Sekiguchi}, {Tatsumi}, \& {Yamamoto}}]{2013PhRvD..88d3007A}
{Aso}, Y., {Michimura}, Y., {Somiya}, K., {et~al.} 2013, \prd, 88, 043007, \dodoi{10.1103/PhysRevD.88.043007}

\bibitem[{{Barausse} {et~al.}(2012){Barausse}, {Morozova}, \& {Rezzolla}}]{2012ApJ...758...63B}
{Barausse}, E., {Morozova}, V., \& {Rezzolla}, L. 2012, \apj, 758, 63, \dodoi{10.1088/0004-637X/758/1/63}

\bibitem[{{Bartos} {et~al.}(2017){Bartos}, {Kocsis}, {Haiman}, \& {M{\'a}rka}}]{2017ApJ...835..165B}
{Bartos}, I., {Kocsis}, B., {Haiman}, Z., \& {M{\'a}rka}, S. 2017, \apj, 835, 165, \dodoi{10.3847/1538-4357/835/2/165}

\bibitem[{{Belczynski} {et~al.}(2016){Belczynski}, {Heger}, {Gladysz}, {Ruiter}, {Woosley}, {Wiktorowicz}, {Chen}, {Bulik}, {O'Shaughnessy}, {Holz}, {Fryer}, \& {Berti}}]{2016A&A...594A..97B}
{Belczynski}, K., {Heger}, A., {Gladysz}, W., {et~al.} 2016, \aap, 594, A97, \dodoi{10.1051/0004-6361/201628980}

\bibitem[{{Bellovary} {et~al.}(2016){Bellovary}, {Mac Low}, {McKernan}, \& {Ford}}]{2016ApJ...819L..17B}
{Bellovary}, J.~M., {Mac Low}, M.-M., {McKernan}, B., \& {Ford}, K.~E.~S. 2016, \apjl, 819, L17, \dodoi{10.3847/2041-8205/819/2/L17}

\bibitem[{{Bogdanovi{\'c}} {et~al.}(2007){Bogdanovi{\'c}}, {Reynolds}, \& {Miller}}]{2007ApJ...661L.147B}
{Bogdanovi{\'c}}, T., {Reynolds}, C.~S., \& {Miller}, M.~C. 2007, \apjl, 661, L147, \dodoi{10.1086/518769}

\bibitem[{{Campanelli} {et~al.}(2007){Campanelli}, {Lousto}, {Zlochower}, \& {Merritt}}]{2007ApJ...659L...5C}
{Campanelli}, M., {Lousto}, C., {Zlochower}, Y., \& {Merritt}, D. 2007, \apjl, 659, L5, \dodoi{10.1086/516712}

\bibitem[{{Chen} \& {Dai}(2024)}]{2024ApJ...961..206C}
{Chen}, K., \& {Dai}, Z.-G. 2024, \apj, 961, 206, \dodoi{10.3847/1538-4357/ad0dfd}

\bibitem[{{Chen} \& {Jani}(2024)}]{2024arXiv241102778C}
{Chen}, S., \& {Jani}, K. 2024, arXiv e-prints, arXiv:2411.02778, \dodoi{10.48550/arXiv.2411.02778}

\bibitem[{{Cheng} {et~al.}(2023){Cheng}, {Zevin}, \& {Vitale}}]{2023ApJ...955..127C}
{Cheng}, A.~Q., {Zevin}, M., \& {Vitale}, S. 2023, \apj, 955, 127, \dodoi{10.3847/1538-4357/aced98}

\bibitem[{{Di Carlo} {et~al.}(2020{\natexlab{a}}){Di Carlo}, {Mapelli}, {Bouffanais}, {Giacobbo}, {Santoliquido}, {Bressan}, {Spera}, \& {Haardt}}]{2020MNRAS.497.1043D}
{Di Carlo}, U.~N., {Mapelli}, M., {Bouffanais}, Y., {et~al.} 2020{\natexlab{a}}, \mnras, 497, 1043, \dodoi{10.1093/mnras/staa1997}

\bibitem[{{Di Carlo} {et~al.}(2020{\natexlab{b}}){Di Carlo}, {Mapelli}, {Giacobbo}, {Spera}, {Bouffanais}, {Rastello}, {Santoliquido}, {Pasquato}, {Ballone}, {Trani}, {Torniamenti}, \& {Haardt}}]{2020MNRAS.498..495D}
{Di Carlo}, U.~N., {Mapelli}, M., {Giacobbo}, N., {et~al.} 2020{\natexlab{b}}, \mnras, 498, 495, \dodoi{10.1093/mnras/staa2286}

\bibitem[{{Di Carlo} {et~al.}(2021){Di Carlo}, {Mapelli}, {Pasquato}, {Rastello}, {Ballone}, {Dall'Amico}, {Giacobbo}, {Iorio}, {Spera}, {Torniamenti}, \& {Haardt}}]{2021MNRAS.507.5132D}
{Di Carlo}, U.~N., {Mapelli}, M., {Pasquato}, M., {et~al.} 2021, \mnras, 507, 5132, \dodoi{10.1093/mnras/stab2390}

\bibitem[{{Di Cintio} {et~al.}(2023){Di Cintio}, {Pasquato}, {Barbieri}, {Trani}, \& {di Carlo}}]{2023A&A...673A...8D}
{Di Cintio}, P., {Pasquato}, M., {Barbieri}, L., {Trani}, A.~A., \& {di Carlo}, U.~N. 2023, \aap, 673, A8, \dodoi{10.1051/0004-6361/202346124}

\bibitem[{{Dominik} {et~al.}(2012){Dominik}, {Belczynski}, {Fryer}, {Holz}, {Berti}, {Bulik}, {Mandel}, \& {O'Shaughnessy}}]{2012ApJ...759...52D}
{Dominik}, M., {Belczynski}, K., {Fryer}, C., {et~al.} 2012, \apj, 759, 52, \dodoi{10.1088/0004-637X/759/1/52}

\bibitem[{{Fishbach} {et~al.}(2017){Fishbach}, {Holz}, \& {Farr}}]{2017ApJ...840L..24F}
{Fishbach}, M., {Holz}, D.~E., \& {Farr}, B. 2017, \apjl, 840, L24, \dodoi{10.3847/2041-8213/aa7045}

\bibitem[{{Ford} \& {McKernan}(2022)}]{2022MNRAS.517.5827F}
{Ford}, K.~E.~S., \& {McKernan}, B. 2022, \mnras, 517, 5827, \dodoi{10.1093/mnras/stac2861}

\bibitem[{{Fragione} {et~al.}(2022){Fragione}, {Kocsis}, {Rasio}, \& {Silk}}]{2022ApJ...927..231F}
{Fragione}, G., {Kocsis}, B., {Rasio}, F.~A., \& {Silk}, J. 2022, \apj, 927, 231, \dodoi{10.3847/1538-4357/ac5026}

\bibitem[{{Gerosa} \& {Berti}(2017)}]{2017PhRvD..95l4046G}
{Gerosa}, D., \& {Berti}, E. 2017, \prd, 95, 124046, \dodoi{10.1103/PhysRevD.95.124046}

\bibitem[{{Gerosa} \& {Berti}(2019)}]{2019PhRvD.100d1301G}
---. 2019, \prd, 100, 041301, \dodoi{10.1103/PhysRevD.100.041301}

\bibitem[{{Gerosa} \& {Fishbach}(2021)}]{2021NatAs...5..749G}
{Gerosa}, D., \& {Fishbach}, M. 2021, Nature Astronomy, 5, 749, \dodoi{10.1038/s41550-021-01398-w}

\bibitem[{{Gonz{\'a}lez Prieto} {et~al.}(2022){Gonz{\'a}lez Prieto}, {Kremer}, {Fragione}, {Martinez}, {Weatherford}, {Zevin}, \& {Rasio}}]{2022ApJ...940..131G}
{Gonz{\'a}lez Prieto}, E., {Kremer}, K., {Fragione}, G., {et~al.} 2022, \apj, 940, 131, \dodoi{10.3847/1538-4357/ac9b0f}

\bibitem[{{Graham} {et~al.}(2020){Graham}, {Ford}, {McKernan}, {Ross}, {Stern}, {Burdge}, {Coughlin}, {Djorgovski}, {Drake}, {Duev}, {Kasliwal}, {Mahabal}, {van Velzen}, {Belecki}, {Bellm}, {Burruss}, {Cenko}, {Cunningham}, {Helou}, {Kulkarni}, {Masci}, {Prince}, {Reiley}, {Rodriguez}, {Rusholme}, {Smith}, \& {Soumagnac}}]{2020PhRvL.124y1102G}
{Graham}, M.~J., {Ford}, K.~E.~S., {McKernan}, B., {et~al.} 2020, \prl, 124, 251102, \dodoi{10.1103/PhysRevLett.124.251102}

\bibitem[{{Graham} {et~al.}(2023){Graham}, {McKernan}, {Ford}, {Stern}, {Djorgovski}, {Coughlin}, {Burdge}, {Bellm}, {Helou}, {Mahabal}, {Masci}, {Purdum}, {Rosnet}, \& {Rusholme}}]{2023ApJ...942...99G}
{Graham}, M.~J., {McKernan}, B., {Ford}, K.~E.~S., {et~al.} 2023, \apj, 942, 99, \dodoi{10.3847/1538-4357/aca480}

\bibitem[{{Heger} {et~al.}(2003){Heger}, {Fryer}, {Woosley}, {Langer}, \& {Hartmann}}]{2003ApJ...591..288H}
{Heger}, A., {Fryer}, C.~L., {Woosley}, S.~E., {Langer}, N., \& {Hartmann}, D.~H. 2003, \apj, 591, 288, \dodoi{10.1086/375341}

\bibitem[{{Heggie}(1975)}]{1975MNRAS.173..729H}
{Heggie}, D.~C. 1975, \mnras, 173, 729, \dodoi{10.1093/mnras/173.3.729}

\bibitem[{{Hills} \& {Fullerton}(1980)}]{1980AJ.....85.1281H}
{Hills}, J.~G., \& {Fullerton}, L.~W. 1980, \aj, 85, 1281, \dodoi{10.1086/112798}

\bibitem[{{Hofmann} {et~al.}(2016){Hofmann}, {Barausse}, \& {Rezzolla}}]{2016ApJ...825L..19H}
{Hofmann}, F., {Barausse}, E., \& {Rezzolla}, L. 2016, \apjl, 825, L19, \dodoi{10.3847/2041-8205/825/2/L19}

\bibitem[{{Iwaya} {et~al.}(2023){Iwaya}, {Kinugawa}, \& {Tagoshi}}]{2023arXiv231217491I}
{Iwaya}, M., {Kinugawa}, T., \& {Tagoshi}, H. 2023, arXiv e-prints, arXiv:2312.17491, \dodoi{10.48550/arXiv.2312.17491}

\bibitem[{{Kesden} {et~al.}(2010){Kesden}, {Sperhake}, \& {Berti}}]{2010ApJ...715.1006K}
{Kesden}, M., {Sperhake}, U., \& {Berti}, E. 2010, \apj, 715, 1006, \dodoi{10.1088/0004-637X/715/2/1006}

\bibitem[{{Kimball} {et~al.}(2021){Kimball}, {Talbot}, {Berry}, {Zevin}, {Thrane}, {Kalogera}, {Buscicchio}, {Carney}, {Dent}, {Middleton}, {Payne}, {Veitch}, \& {Williams}}]{2021ApJ...915L..35K}
{Kimball}, C., {Talbot}, C., {Berry}, C. P.~L., {et~al.} 2021, \apjl, 915, L35, \dodoi{10.3847/2041-8213/ac0aef}

\bibitem[{{Kremer} {et~al.}(2020){Kremer}, {Spera}, {Becker}, {Chatterjee}, {Di Carlo}, {Fragione}, {Rodriguez}, {Ye}, \& {Rasio}}]{2020ApJ...903...45K}
{Kremer}, K., {Spera}, M., {Becker}, D., {et~al.} 2020, \apj, 903, 45, \dodoi{10.3847/1538-4357/abb945}

\bibitem[{{Kritos} {et~al.}(2023){Kritos}, {Berti}, \& {Silk}}]{2023PhRvD.108h3012K}
{Kritos}, K., {Berti}, E., \& {Silk}, J. 2023, \prd, 108, 083012, \dodoi{10.1103/PhysRevD.108.083012}

\bibitem[{{Kroupa}(2001)}]{2001MNRAS.322..231K}
{Kroupa}, P. 2001, \mnras, 322, 231, \dodoi{10.1046/j.1365-8711.2001.04022.x}

\bibitem[{{Leong} {et~al.}(2024){Leong}, {Janquart}, {Sharma}, {Martens}, {Ajith}, \& {Hannuksela}}]{2024arXiv240813144L}
{Leong}, S. H.~W., {Janquart}, J., {Sharma}, A.~K., {et~al.} 2024, arXiv e-prints, arXiv:2408.13144, \dodoi{10.48550/arXiv.2408.13144}

\bibitem[{{Li}(2022{\natexlab{a}})}]{2022PhRvD.105f3006L}
{Li}, G.-P. 2022{\natexlab{a}}, \prd, 105, 063006, \dodoi{10.1103/PhysRevD.105.063006}

\bibitem[{{Li}(2022{\natexlab{b}})}]{2022A&A...666A.194L}
---. 2022{\natexlab{b}}, \aap, 666, A194, \dodoi{10.1051/0004-6361/202244257}

\bibitem[{{Li} {et~al.}(2023{\natexlab{a}}){Li}, {Lin}, \& {Yuan}}]{2023PhRvD.107f3007L}
{Li}, G.-P., {Lin}, D.-B., \& {Yuan}, Y. 2023{\natexlab{a}}, \prd, 107, 063007, \dodoi{10.1103/PhysRevD.107.063007}

\bibitem[{{Li} {et~al.}(2023{\natexlab{b}}){Li}, {Dempsey}, {Li}, {Lai}, \& {Li}}]{2023ApJ...944L..42L}
{Li}, J., {Dempsey}, A.~M., {Li}, H., {Lai}, D., \& {Li}, S. 2023{\natexlab{b}}, \apjl, 944, L42, \dodoi{10.3847/2041-8213/acb934}

\bibitem[{{Li} {et~al.}(2024{\natexlab{a}}){Li}, {Tang}, {Gao}, {Wu}, \& {Wang}}]{2024ApJ...977...67L}
{Li}, Y.-J., {Tang}, S.-P., {Gao}, S.-J., {Wu}, D.-C., \& {Wang}, Y.-Z. 2024{\natexlab{a}}, \apj, 977, 67, \dodoi{10.3847/1538-4357/ad83b5}

\bibitem[{{Li} {et~al.}(2024{\natexlab{b}}){Li}, {Wang}, {Tang}, \& {Fan}}]{2024PhRvL.133e1401L}
{Li}, Y.-J., {Wang}, Y.-Z., {Tang}, S.-P., \& {Fan}, Y.-Z. 2024{\natexlab{b}}, \prl, 133, 051401, \dodoi{10.1103/PhysRevLett.133.051401}

\bibitem[{{LIGO Scientific Collaboration} {et~al.}(2015){LIGO Scientific Collaboration}, {Aasi}, {Abbott}, {Abbott}, {Abbott}, {Abernathy}, {Ackley}, {Adams}, {Adams}, {Addesso}, \& et~al.}]{2015CQGra..32g4001L}
{LIGO Scientific Collaboration}, {Aasi}, J., {Abbott}, B.~P., {et~al.} 2015, Classical and Quantum Gravity, 32, 074001, \dodoi{10.1088/0264-9381/32/7/074001}

\bibitem[{{Liu} {et~al.}(2024){Liu}, {Hartwig}, {Sartorio}, {Dvorkin}, {Costa}, {Santoliquido}, {Fialkov}, {Klessen}, \& {Bromm}}]{2024MNRAS.534.1634L}
{Liu}, B., {Hartwig}, T., {Sartorio}, N.~S., {et~al.} 2024, \mnras, 534, 1634, \dodoi{10.1093/mnras/stae2120}

\bibitem[{{Luo} {et~al.}(2023){Luo}, {Wu}, {Zhang}, {Wang}, {Ho}, \& {Yuan}}]{2023MNRAS.524.6015L}
{Luo}, Y., {Wu}, X.-J., {Zhang}, S.-R., {et~al.} 2023, \mnras, 524, 6015, \dodoi{10.1093/mnras/stad2188}

\bibitem[{{Mahapatra} {et~al.}(2024){Mahapatra}, {Chattopadhyay}, {Gupta}, {Antonini}, {Favata}, {Sathyaprakash}, \& {Arun}}]{2024ApJ...975..117M}
{Mahapatra}, P., {Chattopadhyay}, D., {Gupta}, A., {et~al.} 2024, \apj, 975, 117, \dodoi{10.3847/1538-4357/ad781b}

\bibitem[{{Mahapatra} {et~al.}(2022){Mahapatra}, {Chattopadhyay}, {Gupta}, {Favata}, {Sathyaprakash}, \& {Arun}}]{2022arXiv220905766M}
---. 2022, arXiv e-prints, arXiv:2209.05766, \dodoi{10.48550/arXiv.2209.05766}

\bibitem[{{Mahapatra} {et~al.}(2021){Mahapatra}, {Gupta}, {Favata}, {Arun}, \& {Sathyaprakash}}]{2021ApJ...918L..31M}
{Mahapatra}, P., {Gupta}, A., {Favata}, M., {Arun}, K.~G., \& {Sathyaprakash}, B.~S. 2021, \apjl, 918, L31, \dodoi{10.3847/2041-8213/ac20db}

\bibitem[{{Mandel} \& {Broekgaarden}(2022)}]{2022LRR....25....1M}
{Mandel}, I., \& {Broekgaarden}, F.~S. 2022, Living Reviews in Relativity, 25, 1, \dodoi{10.1007/s41114-021-00034-3}

\bibitem[{{Mandel} \& {Farmer}(2022)}]{2022PhR...955....1M}
{Mandel}, I., \& {Farmer}, A. 2022, \physrep, 955, 1, \dodoi{10.1016/j.physrep.2022.01.003}

\bibitem[{{Mandel} {et~al.}(2019){Mandel}, {Farr}, \& {Gair}}]{2019MNRAS.486.1086M}
{Mandel}, I., {Farr}, W.~M., \& {Gair}, J.~R. 2019, \mnras, 486, 1086, \dodoi{10.1093/mnras/stz896}

\bibitem[{{Mapelli}(2021)}]{2021hgwa.bookE..16M}
{Mapelli}, M. 2021, in Handbook of Gravitational Wave Astronomy, 16, \dodoi{10.1007/978-981-15-4702-7_16-1}

\bibitem[{{Mapelli} {et~al.}(2021){Mapelli}, {Dall'Amico}, {Bouffanais}, {Giacobbo}, {Arca Sedda}, {Artale}, {Ballone}, {Di Carlo}, {Iorio}, {Santoliquido}, \& {Torniamenti}}]{2021MNRAS.505..339M}
{Mapelli}, M., {Dall'Amico}, M., {Bouffanais}, Y., {et~al.} 2021, \mnras, 505, 339, \dodoi{10.1093/mnras/stab1334}

\bibitem[{{McKernan} {et~al.}(2012){McKernan}, {Ford}, {Lyra}, \& {Perets}}]{2012MNRAS.425..460M}
{McKernan}, B., {Ford}, K.~E.~S., {Lyra}, W., \& {Perets}, H.~B. 2012, \mnras, 425, 460, \dodoi{10.1111/j.1365-2966.2012.21486.x}

\bibitem[{{McKernan} {et~al.}(2020){McKernan}, {Ford}, {O'Shaugnessy}, \& {Wysocki}}]{2020MNRAS.494.1203M}
{McKernan}, B., {Ford}, K.~E.~S., {O'Shaugnessy}, R., \& {Wysocki}, D. 2020, \mnras, 494, 1203, \dodoi{10.1093/mnras/staa740}

\bibitem[{{McKernan} {et~al.}(2019){McKernan}, {Ford}, {Bartos}, {Graham}, {Lyra}, {Marka}, {Marka}, {Ross}, {Stern}, \& {Yang}}]{2019ApJ...884L..50M}
{McKernan}, B., {Ford}, K.~E.~S., {Bartos}, I., {et~al.} 2019, \apjl, 884, L50, \dodoi{10.3847/2041-8213/ab4886}

\bibitem[{{Morscher} {et~al.}(2015){Morscher}, {Pattabiraman}, {Rodriguez}, {Rasio}, \& {Umbreit}}]{2015ApJ...800....9M}
{Morscher}, M., {Pattabiraman}, B., {Rodriguez}, C., {Rasio}, F.~A., \& {Umbreit}, S. 2015, \apj, 800, 9, \dodoi{10.1088/0004-637X/800/1/9}

\bibitem[{{Mould} {et~al.}(2022){Mould}, {Gerosa}, \& {Taylor}}]{2022PhRvD.106j3013M}
{Mould}, M., {Gerosa}, D., \& {Taylor}, S.~R. 2022, \prd, 106, 103013, \dodoi{10.1103/PhysRevD.106.103013}

\bibitem[{{Nitz} {et~al.}(2023){Nitz}, {Kumar}, {Wang}, {Kastha}, {Wu}, {Sch{\"a}fer}, {Dhurkunde}, \& {Capano}}]{2023ApJ...946...59N}
{Nitz}, A.~H., {Kumar}, S., {Wang}, Y.-F., {et~al.} 2023, \apj, 946, 59, \dodoi{10.3847/1538-4357/aca591}

\bibitem[{{O'Leary} {et~al.}(2016){O'Leary}, {Meiron}, \& {Kocsis}}]{2016ApJ...824L..12O}
{O'Leary}, R.~M., {Meiron}, Y., \& {Kocsis}, B. 2016, \apjl, 824, L12, \dodoi{10.3847/2041-8205/824/1/L12}

\bibitem[{{Pavl{\'\i}k} \& {Vesperini}(2022)}]{2022MNRAS.515.1830P}
{Pavl{\'\i}k}, V., \& {Vesperini}, E. 2022, \mnras, 515, 1830, \dodoi{10.1093/mnras/stac1776}

\bibitem[{{Punturo} {et~al.}(2010{\natexlab{a}}){Punturo}, {Abernathy}, {Acernese}, {Allen}, {Andersson}, {Arun}, {Barone}, {Barr}, {Barsuglia}, {Beker}, {Beveridge}, {Birindelli}, {Bose}, {Bosi}, {Braccini}, {Bradaschia}, {Bulik}, {Calloni}, {Cella}, {Chassande Mottin}, {Chelkowski}, {Chincarini}, {Clark}, {Coccia}, {Colacino}, {Colas}, {Cumming}, {Cunningham}, {Cuoco}, {Danilishin}, {Danzmann}, {De Luca}, {De Salvo}, {Dent}, {De Rosa}, {Di Fiore}, {Di Virgilio}, {Doets}, {Fafone}, {Falferi}, {Flaminio}, {Franc}, {Frasconi}, {Freise}, {Fulda}, {Gair}, {Gemme}, {Gennai}, {Giazotto}, {Glampedakis}, {Granata}, {Grote}, {Guidi}, {Hammond}, {Hannam}, {Harms}, {Heinert}, {Hendry}, {Heng}, {Hennes}, {Hild}, {Hough}, {Husa}, {Huttner}, {Jones}, {Khalili}, {Kokeyama}, {Kokkotas}, {Krishnan}, {Lorenzini}, {L{\"u}ck}, {Majorana}, {Mandel}, {Mandic}, {Martin}, {Michel}, {Minenkov}, {Morgado}, {Mosca}, {Mours}, {M{\"u}ller{\textendash}Ebhardt}, {Murray}, {Nawrodt}, {Nelson}, {Oshaughnessy}, {Ott}, {Palomba}, {Paoli},
  {Parguez}, {Pasqualetti}, {Passaquieti}, {Passuello}, {Pinard}, {Poggiani}, {Popolizio}, {Prato}, {Puppo}, {Rabeling}, {Rapagnani}, {Read}, {Regimbau}, {Rehbein}, {Reid}, {Rezzolla}, {Ricci}, {Richard}, {Rocchi}, {Rowan}, {R{\"u}diger}, {Sassolas}, {Sathyaprakash}, {Schnabel}, {Schwarz}, {Seidel}, {Sintes}, {Somiya}, {Speirits}, {Strain}, {Strigin}, {Sutton}, {Tarabrin}, {Th{\"u}ring}, {van den Brand}, {van Leewen}, {van Veggel}, {van den Broeck}, {Vecchio}, {Veitch}, {Vetrano}, {Vicere}, {Vyatchanin}, {Willke}, {Woan}, {Wolfango}, \& {Yamamoto}}]{2010CQGra..27s4002P}
{Punturo}, M., {Abernathy}, M., {Acernese}, F., {et~al.} 2010{\natexlab{a}}, Classical and Quantum Gravity, 27, 194002, \dodoi{10.1088/0264-9381/27/19/194002}

\bibitem[{{Punturo} {et~al.}(2010{\natexlab{b}}){Punturo}, {Abernathy}, {Acernese}, {Allen}, {Andersson}, {Arun}, {Barone}, {Barr}, {Barsuglia}, {Beker}, {Beveridge}, {Birindelli}, {Bose}, {Bosi}, {Braccini}, {Bradaschia}, {Bulik}, {Calloni}, {Cella}, {Chassande Mottin}, {Chelkowski}, {Chincarini}, {Clark}, {Coccia}, {Colacino}, {Colas}, {Cumming}, {Cunningham}, {Cuoco}, {Danilishin}, {Danzmann}, {De Luca}, {De Salvo}, {Dent}, {Derosa}, {Di Fiore}, {Di Virgilio}, {Doets}, {Fafone}, {Falferi}, {Flaminio}, {Franc}, {Frasconi}, {Freise}, {Fulda}, {Gair}, {Gemme}, {Gennai}, {Giazotto}, {Glampedakis}, {Granata}, {Grote}, {Guidi}, {Hammond}, {Hannam}, {Harms}, {Heinert}, {Hendry}, {Heng}, {Hennes}, {Hild}, {Hough}, {Husa}, {Huttner}, {Jones}, {Khalili}, {Kokeyama}, {Kokkotas}, {Krishnan}, {Lorenzini}, {L{\"u}ck}, {Majorana}, {Mandel}, {Mandic}, {Martin}, {Michel}, {Minenkov}, {Morgado}, {Mosca}, {Mours}, {M{\"u}ller-Ebhardt}, {Murray}, {Nawrodt}, {Nelson}, {Oshaughnessy}, {Ott}, {Palomba}, {Paoli}, {Parguez},
  {Pasqualetti}, {Passaquieti}, {Passuello}, {Pinard}, {Poggiani}, {Popolizio}, {Prato}, {Puppo}, {Rabeling}, {Rapagnani}, {Read}, {Regimbau}, {Rehbein}, {Reid}, {Rezzolla}, {Ricci}, {Richard}, {Rocchi}, {Rowan}, {R{\"u}diger}, {Sassolas}, {Sathyaprakash}, {Schnabel}, {Schwarz}, {Seidel}, {Sintes}, {Somiya}, {Speirits}, {Strain}, {Strigin}, {Sutton}, {Tarabrin}, {van den Brand}, {van Leewen}, {van Veggel}, {van den Broeck}, {Vecchio}, {Veitch}, {Vetrano}, {Vicere}, {Vyatchanin}, {Willke}, {Woan}, {Wolfango}, \& {Yamamoto}}]{2010CQGra..27h4007P}
---. 2010{\natexlab{b}}, Classical and Quantum Gravity, 27, 084007, \dodoi{10.1088/0264-9381/27/8/084007}

\bibitem[{{Rizzuto} {et~al.}(2022){Rizzuto}, {Naab}, {Spurzem}, {Arca-Sedda}, {Giersz}, {Ostriker}, \& {Banerjee}}]{2022MNRAS.512..884R}
{Rizzuto}, F.~P., {Naab}, T., {Spurzem}, R., {et~al.} 2022, \mnras, 512, 884, \dodoi{10.1093/mnras/stac231}

\bibitem[{{Rizzuto} {et~al.}(2021){Rizzuto}, {Naab}, {Spurzem}, {Giersz}, {Ostriker}, {Stone}, {Wang}, {Berczik}, \& {Rampp}}]{2021MNRAS.501.5257R}
---. 2021, \mnras, 501, 5257, \dodoi{10.1093/mnras/staa3634}

\bibitem[{{Rodriguez} {et~al.}(2019){Rodriguez}, {Zevin}, {Amaro-Seoane}, {Chatterjee}, {Kremer}, {Rasio}, \& {Ye}}]{2019PhRvD.100d3027R}
{Rodriguez}, C.~L., {Zevin}, M., {Amaro-Seoane}, P., {et~al.} 2019, \prd, 100, 043027, \dodoi{10.1103/PhysRevD.100.043027}

\bibitem[{{Somiya}(2012)}]{2012CQGra..29l4007S}
{Somiya}, K. 2012, Classical and Quantum Gravity, 29, 124007, \dodoi{10.1088/0264-9381/29/12/124007}

\bibitem[{{Tagawa} {et~al.}(2020{\natexlab{a}}){Tagawa}, {Haiman}, {Bartos}, \& {Kocsis}}]{2020ApJ...899...26T}
{Tagawa}, H., {Haiman}, Z., {Bartos}, I., \& {Kocsis}, B. 2020{\natexlab{a}}, \apj, 899, 26, \dodoi{10.3847/1538-4357/aba2cc}

\bibitem[{{Tagawa} {et~al.}(2020{\natexlab{b}}){Tagawa}, {Haiman}, \& {Kocsis}}]{2020ApJ...898...25T}
{Tagawa}, H., {Haiman}, Z., \& {Kocsis}, B. 2020{\natexlab{b}}, \apj, 898, 25, \dodoi{10.3847/1538-4357/ab9b8c}

\bibitem[{{Tagawa} {et~al.}(2023){Tagawa}, {Kimura}, \& {Haiman}}]{2023ApJ...955...23T}
{Tagawa}, H., {Kimura}, S.~S., \& {Haiman}, Z. 2023, \apj, 955, 23, \dodoi{10.3847/1538-4357/ace71d}

\bibitem[{{Tagawa} {et~al.}(2021){Tagawa}, {Kocsis}, {Haiman}, {Bartos}, {Omukai}, \& {Samsing}}]{2021ApJ...908..194T}
{Tagawa}, H., {Kocsis}, B., {Haiman}, Z., {et~al.} 2021, \apj, 908, 194, \dodoi{10.3847/1538-4357/abd555}

\bibitem[{{The LIGO Scientific Collaboration} {et~al.}(2021){The LIGO Scientific Collaboration}, {the Virgo Collaboration}, {Abbott}, {Abbott}, {Acernese}, {Ackley}, {Adams}, {Adhikari}, {Adhikari}, {Adya}, \& et~al.}]{2021arXiv210801045T}
{The LIGO Scientific Collaboration}, {the Virgo Collaboration}, {Abbott}, R., {et~al.} 2021, arXiv e-prints, arXiv:2108.01045, \dodoi{10.48550/arXiv.2108.01045}

\bibitem[{{Thrane} \& {Talbot}(2019)}]{2019PASA...36...10T}
{Thrane}, E., \& {Talbot}, C. 2019, \pasa, 36, e010, \dodoi{10.1017/pasa.2019.2}

\bibitem[{{Vaccaro} {et~al.}(2024){Vaccaro}, {Mapelli}, {P{\'e}rigois}, {Barone}, {Artale}, {Dall'Amico}, {Iorio}, \& {Torniamenti}}]{2024A&A...685A..51V}
{Vaccaro}, M.~P., {Mapelli}, M., {P{\'e}rigois}, C., {et~al.} 2024, \aap, 685, A51, \dodoi{10.1051/0004-6361/202348509}

\bibitem[{{Veronesi} {et~al.}(2023){Veronesi}, {Rossi}, \& {van Velzen}}]{2023MNRAS.526.6031V}
{Veronesi}, N., {Rossi}, E.~M., \& {van Velzen}, S. 2023, \mnras, 526, 6031, \dodoi{10.1093/mnras/stad3157}

\bibitem[{{Veronesi} {et~al.}(2022){Veronesi}, {Rossi}, {van Velzen}, \& {Buscicchio}}]{2022MNRAS.514.2092V}
{Veronesi}, N., {Rossi}, E.~M., {van Velzen}, S., \& {Buscicchio}, R. 2022, \mnras, 514, 2092, \dodoi{10.1093/mnras/stac1346}

\bibitem[{{Veronesi} {et~al.}(2024{\natexlab{a}}){Veronesi}, {van Velzen}, \& {Rossi}}]{2024arXiv240505318V}
{Veronesi}, N., {van Velzen}, S., \& {Rossi}, E.~M. 2024{\natexlab{a}}, arXiv e-prints, arXiv:2405.05318, \dodoi{10.48550/arXiv.2405.05318}

\bibitem[{{Veronesi} {et~al.}(2024{\natexlab{b}}){Veronesi}, {van Velzen}, {Rossi}, \& {Storey-Fisher}}]{2024arXiv240721568V}
{Veronesi}, N., {van Velzen}, S., {Rossi}, E.~M., \& {Storey-Fisher}, K. 2024{\natexlab{b}}, arXiv e-prints, arXiv:2407.21568, \dodoi{10.48550/arXiv.2407.21568}

\bibitem[{{Wang} {et~al.}(2021){Wang}, {Liu}, {Ho}, {Li}, \& {Du}}]{2021ApJ...916L..17W}
{Wang}, J.-M., {Liu}, J.-R., {Ho}, L.~C., {Li}, Y.-R., \& {Du}, P. 2021, \apjl, 916, L17, \dodoi{10.3847/2041-8213/ac0b46}

\bibitem[{{Wang} {et~al.}(2022){Wang}, {Li}, {Vink}, {Fan}, {Tang}, {Qin}, \& {Wei}}]{2022ApJ...941L..39W}
{Wang}, Y.-Z., {Li}, Y.-J., {Vink}, J.~S., {et~al.} 2022, \apjl, 941, L39, \dodoi{10.3847/2041-8213/aca89f}

\bibitem[{{Woosley} {et~al.}(2007){Woosley}, {Blinnikov}, \& {Heger}}]{2007Natur.450..390W}
{Woosley}, S.~E., {Blinnikov}, S., \& {Heger}, A. 2007, \nat, 450, 390, \dodoi{10.1038/nature06333}

\bibitem[{{Yang} {et~al.}(2019{\natexlab{a}}){Yang}, {Bartos}, {Haiman}, {Kocsis}, {M{\'a}rka}, {Stone}, \& {M{\'a}rka}}]{2019ApJ...876..122Y}
{Yang}, Y., {Bartos}, I., {Haiman}, Z., {et~al.} 2019{\natexlab{a}}, \apj, 876, 122, \dodoi{10.3847/1538-4357/ab16e3}

\bibitem[{{Yang} {et~al.}(2020){Yang}, {Gayathri}, {Bartos}, {Haiman}, {Safarzadeh}, \& {Tagawa}}]{2020ApJ...901L..34Y}
{Yang}, Y., {Gayathri}, V., {Bartos}, I., {et~al.} 2020, \apjl, 901, L34, \dodoi{10.3847/2041-8213/abb940}

\bibitem[{{Yang} {et~al.}(2019{\natexlab{b}}){Yang}, {Bartos}, {Gayathri}, {Ford}, {Haiman}, {Klimenko}, {Kocsis}, {M{\'a}rka}, {M{\'a}rka}, {McKernan}, \& {O'Shaughnessy}}]{2019PhRvL.123r1101Y}
{Yang}, Y., {Bartos}, I., {Gayathri}, V., {et~al.} 2019{\natexlab{b}}, \prl, 123, 181101, \dodoi{10.1103/PhysRevLett.123.181101}

\bibitem[{{Yi} \& {Cheng}(2019)}]{2019ApJ...884L..12Y}
{Yi}, S.-X., \& {Cheng}, K.~S. 2019, \apjl, 884, L12, \dodoi{10.3847/2041-8213/ab459a}

\bibitem[{{Zevin} \& {Holz}(2022)}]{2022ApJ...935L..20Z}
{Zevin}, M., \& {Holz}, D.~E. 2022, \apjl, 935, L20, \dodoi{10.3847/2041-8213/ac853d}

\bibitem[{{Zevin} {et~al.}(2021){Zevin}, {Bavera}, {Berry}, {Kalogera}, {Fragos}, {Marchant}, {Rodriguez}, {Antonini}, {Holz}, \& {Pankow}}]{2021ApJ...910..152Z}
{Zevin}, M., {Bavera}, S.~S., {Berry}, C. P.~L., {et~al.} 2021, \apj, 910, 152, \dodoi{10.3847/1538-4357/abe40e}

\bibitem[{{Zhang} {et~al.}(2024){Zhang}, {Zhu}, \& {Yu}}]{2024arXiv240610904Z}
{Zhang}, H.-H., {Zhu}, J.-P., \& {Yu}, Y.-W. 2024, arXiv e-prints, arXiv:2406.10904, \dodoi{10.48550/arXiv.2406.10904}

\bibitem[{{Zhou} {et~al.}(2023){Zhou}, {Zhu}, \& {Wang}}]{2023ApJ...951...74Z}
{Zhou}, Z.-H., {Zhu}, J.-P., \& {Wang}, K. 2023, \apj, 951, 74, \dodoi{10.3847/1538-4357/acd380}

\bibitem[{{Zhu}(2024)}]{2024MNRAS.528L..88Z}
{Zhu}, J.-P. 2024, \mnras, 528, L88, \dodoi{10.1093/mnrasl/slad176}

\end{thebibliography}

\end{document}